# Interpretable Survival Prediction for Colorectal Cancer using Deep Learning

## Authors

Ellery Wulczyn[1], David F. Steiner[1], Melissa Moran[1], Markus Plass[2], Robert Reihs[2], Fraser Tan[1], Isabelle Flament-Auvigne[3], Trissia Brown[3], Peter Regitnig[2], Po-Hsuan Cameron Chen[1], Narayan Hegde[1], Apaar Sadhwani[1], Robert MacDonald[1], Benny Ayalew[1], Greg S. Corrado[1], Lily H. Peng[1], Daniel Tse[1], Heimo Müller[2], Zhaoyang Xu[1], Yun Liu[1], Martin C. Stumpe[4], Kurt Zatloukal[2,†], Craig H. Mermel[1,†]

[1]Google Health, Palo Alto, CA, USA
[2]Medical University of Graz, Graz, Austria
[3]Work done at Google Health via Advanced Clinical
[4]Work done at Google Health. Present address: Tempus Labs Inc, Chicago, United States
[†]Equal contribution

Address correspondence to: Y.L. (liuyun@google.com), K.Z. (kurt.zatloukal@medunigraz.at)

# Abstract

Deriving interpretable prognostic features from deep-learning-based prognostic histopathology models remains a challenge. In this study, we developed a deep learning system (DLS) for predicting disease specific survival for stage II and III colorectal cancer using 3,652 cases (27,300 slides). When evaluated on two validation datasets containing 1,239 cases (9,340 slides) and 738 cases (7,140 slides) respectively, the DLS achieved a 5-year disease-specific survival AUC of 0.70 (95%CI 0.66-0.73) and 0.69 (95%CI 0.64-0.72), and added significant predictive value to a set of 9 clinicopathologic features. To interpret the DLS, we explored the ability of different human-interpretable features to explain the variance in DLS scores. We observed that clinicopathologic features such as T-category, N-category, and grade explained a small fraction of the variance in DLS scores ($R^2$=18% in both validation sets). Next, we generated human-interpretable histologic features by clustering embeddings from a deep-learning based image-similarity model and showed that they explain the majority of the variance ($R^2$ of 73% to 80%). Furthermore, the clustering-derived feature most strongly associated with high DLS scores was also highly prognostic in isolation. With a distinct visual appearance (poorly differentiated tumor cell clusters adjacent to adipose tissue), this feature was identified by annotators with 87.0-95.5% accuracy. Our approach can be used to explain predictions from a prognostic deep learning model and uncover potentially-novel prognostic features that can be reliably identified by people for future validation studies.

# Introduction

Understanding and characterizing a patient's cancer in order to estimate prognosis is essential for treatment decisions. Cancer staging systems, such as TNM classification, were created to categorize patients into different groups with distinct outcomes [1]. However, even within a specific TNM stage there is often substantial variability in patient outcomes. While additional data such as clinical variables, histopathologic parameters, and molecular features can provide important information [2,3], there remains a need for more precise patient risk stratification to improve patient management and disease outcomes. In recent years, there has been a surge of interest in developing machine learning methods to provide novel prognostic information that is not captured in current staging guidelines [4–8]. However, despite some existing efforts to understand machine-learned prognostic features, strategies to gain insights into such features remain limited. If the learned features can be reproducibly identified and demonstrated to have independent prognostic value, this could enable discovery of potentially novel features as well as build the necessary trust for AI-supported decision making in medicine.

A specific use case of the role of prognostication in guiding treatment decisions can be found with colorectal adenocarcinoma, which is the third-most commonly diagnosed cancer and second only to lung cancer in terms of cancer mortality [9]. For stage II patients, adjuvant chemotherapy can be beneficial following resection of the tumor for a small subset of patients, but identifying the high risk patients most likely to benefit represents a clinical challenge as overtreatment can result in substantial adverse effects [10,11]. For patients with stage III disease, although adjuvant chemotherapy is generally the standard of care, prognostic information has important implications for therapy regimen and duration [12]. Known histoprognostic features such as tumor budding and lymphovascular invasion among others can provide useful information, but challenges in both sensitivity and inter-pathologist variability limit their utility [2,13–15]. Better risk stratification within stage II and stage III colorectal cancer therefore offers opportunities to improve therapy decisions and patient care.

Previous machine learning-based efforts to predict clinical outcome using histopathology samples have used one of two main approaches [16]. The first strategy focuses on extraction of pre-defined morphologic features using custom tools such as CellProfiler [17,18], followed by statistical or machine learning techniques to understand which of the pre-defined features are correlated with survival [5,7,8,19,20]. The second and more recent strategy involves use of weakly-supervised deep learning approaches to directly predict survival from whole slide images [4,6,21,22], thus eliminating reliance on pre-defined features but introducing additional challenges in regards to model explainability. While some weakly-supervised studies have tried to visualize the morphological features learned by the models [21,23,24], providing reproducible descriptions of such features and evaluating the extent to which they actually explain the model predictions remain as challenges. In this study, we first present a weakly-supervised deep learning system (DLS) for predicting disease-specific survival (DSS) in colorectal cancer patients and then develop a method for generating human-interpretable histologic features that can both explain the DLS predictions and be used as independent prognostic features.

# Results

**Data Cohorts**

This study included two cohorts of colorectal cancer cases. The first cohort spanned the years from 1984-2007. It was randomly split into a development set of 3,652 cases (which was further split into training and tuning sets, see Methods) and a held-out validation set of 1,239 cases (validation set 1). A second cohort of 738 colorectal cancer cases from 2008-2013 served as a second held-out validation set (validation set 2) to evaluate temporal generalization of the model to a more recent cohort (Table 1, Supplementary Figure S1). Patient characteristics of the two validation sets are reported in Supplementary Table S1.

**Tumor Segmentation Model**

We first developed a tumor segmentation model for the purpose of categorizing every region on a whole-slide image as tumor or non-tumor. This model was developed using pixel-level annotations provided for a subset of slides from the overall training split (Supplementary Figure S1) and was evaluated on a held-out set of slides, also from the overall training split (44 slides, 6,866,573 patches, Supplementary Figures S2-S4). For classifying individual image patches as tumor vs. non-tumor, this model achieved an area under the receiver operating characteristic curve (AUC) of 0.985 (95%CI 0.984-0.985). Using this model to identify regions of interest for the prognostic model instead of a simple tissue detector substantially improved the performance of the prognostic model (Methods, Supplementary Figure S5).

**Evaluating DLS Performance**

The regions identified by the tumor segmentation model were used as the input for a second, prognostic model to produce case-level risk scores. The tumor segmentation model and prognostic model were applied sequentially to predict prognosis for each case, and are collectively referred to as the DLS.

We evaluated the ability of the DLS to predict DSS in two separate held-out validation sets (each comprising cases from different time periods). Validation set 1 had 10-35 years of follow-up, while the cases in the more recent validation set 2 had 5-9 years of follow-up. Thus, to allow direct comparisons across the two validation sets, we used the AUC for 5-year DSS, which is not affected by the differences in follow up period available for the two validation sets. For stage II cases, the DLS demonstrated a 5-year AUC of 0.680 in validation set 1 and 0.663 in validation set 2 (Table 2). The 5-year AUC for stage III cases was 0.655 in both validation sets. In the combined cohorts of stage II and stage III cases, the 5-year AUC was 0.698 and 0.686 for the two validation sets, respectively. The 95% confidence intervals (CIs) are provided in Table 2.

In Kaplan-Meier analysis, the DLS demonstrated significant risk stratification in both validation sets ($p<0.001$ for log-rank test comparing the high and low risk DLS prediction quartiles; Figure 1A). The 5-year DSS rates of the high and low risk groups among stage II cases were 64% and 89% respectively in validation set 1. In validation set 2, the difference in survival rates between risk groups was similar with 5-year DSS of 60% (high risk) vs. 86% (low risk). For stage III cases, the survival rates for the high and low risk groups were 35% versus 66% in validation set 1 and 42% versus 74% in validation set 2. Similar results were observed for analysis over the combined cohort of stage II/III cases (Supplementary Table S2).

We further performed univariable and multivariable Cox regressions for both the DLS and clinicopathologic features (age, sex, tumor grade, and T, N, R, L, and V categories). The univariable analysis showed that the DLS was significantly associated with DSS for both stage II and stage III as well for the combined stage II/III cohort in both validation sets (p<0.001; Supplementary Table S3). After adjusting for the clinicopathologic features, the DLS remained a significant predictor of DSS (p<0.001; Table 3). We also compared the 5-year AUC of the Cox models containing the clinicopathologic features to those that additionally incorporated the DLS-assigned risk score (Supplementary Table S4A). For stage II, addition of the DLS to the clinicopathologic features increased 5-year AUC over the clinicopathologic features alone by 0.120 and 0.085 for the two validation sets. For stage III, the corresponding increase over the clinicopathologic features alone was 0.065 (validation set 1) and 0.022 (validation set 2). For the combined stage II/III cases, the absolute increases were 0.055 and 0.038 with final AUCs of 0.733 and 0.721, respectively. The increases in prognostic value provided by the addition of the DLS were also observed based on c-index analysis (Supplementary Table S5). Finally, to more directly address the possibility of DLS correlation with depth of tumor invasion, we performed subanalysis on the T3 cases only. The performance of the DLS remained similar for this T3 subanalysis (Supplementary Table S6A).

**Understanding DLS Predictions**

Because the DLS was developed in a weakly-supervised fashion without specifically being trained to predict known clinicopathologic features, we sought to understand what features were most highly associated with the DLS predictions. Specifically, we fit regression models to predict DLS scores using both the set of clinicopathologic features described above and a set of clustering-derived features (described below). Regression coefficients for individual features were used to evaluate the association between the DLS and individual features, while the adjusted coefficient of determination ($R^2$) was used to measure the fraction of variance in DLS scores explained by each feature set.

**DLS Association with Clinicopathologic Features**

We first examined the association of the DLS with clinicopathologic features (Table 4A). The features most significantly associated with the DLS risk score were the T and N categories. Specifically, cases with higher T and N categories also had higher DLS risk scores. Similar observations were made in a univariable correlation analysis (Supplementary Table S7A). Overall, the clinicopathologic features had an $R^2$ of 0.18 (ie, they explained only 18% of the variance in the DLS scores) in both validation sets, indicating that these clinicopathologic features leave a substantial proportion of the variance in DLS scores unexplained.

**DLS Association with Clustering-Derived Features**

Next, given the limited ability of existing clinicopathologic features to explain the variance in DLS scores, we generated a set of 200 human-interpretable histologic features by clustering embeddings from a deep-learning based image-similarity model [25,26]. We then quantified the variance in DLS scores explained by the case-level quantitation of these clustering-derived features (as done above for clinicopathologic features). All 200 features combined demonstrated an $R^2$ of 0.73 for validation set 1 and an $R^2$ of 0.80 for validation set 2 (Table 4B). A subset of 10 of these features selected via forward stepwise selection achieved an $R^2$ of 0.57 for validation set 1 and an $R^2$ of 0.61 for validation set 2.

For each of these top 10 features, sample image patches exhibiting the feature (Figure 2) were formally reviewed by three pathologists (Table 4B). The feature with the highest regression coefficient was characterized by small, moderately-to-poorly differentiated tumor cell clusters adjacent to a substantial component of adipose tissue (cluster #72, Figure 2, and Figure 3A). In the remainder of this paper, we will reference this particular feature as tumor-adipose feature (TAF). Another cluster with a high coefficient (cluster 139) was notable for predominant stroma consisting of intermediate and a mature desmoplastic reaction with a relatively small amount of low-to-intermediate grade tumor. In general, the features associated with higher risk DLS predictions involved intermediate to high grade tumor in small or solid clusters while the lower risk feature clusters typically contained lower grade tumor forming glands and tubules and with high tumor to stroma ratio (Table 4B, Figures 2 and 3A). No remarkable findings were observed in regards to desmoplasia or tumor infiltrating lymphocytes (TILs) across these 10 feature clusters.

**DLS Association with Patch-Level Histoprognostic Features**

The analyses above were performed for case-level DLS scores and case-level quantitation of the clustering-derived features. To gain further insight into the DLS, we compared the average patch-level DLS score for a set of known histoprognostic features as well as the top 10 clustering-derived features (Table 5). Known histoprognostic features were annotated by pathologists on a subset of validation set slides in order to provide patches for analysis (Methods). Among the known features, patches with lymphovascular invasion and perineural invasion had the highest average DLS scores (1.03 and 0.75, respectively), while patches from polyps had the lowest average score (-0.86). Among the top 10 clustering-derived features, the TAF patches had the highest average score (2.76) substantially higher than the other 3 high risk features identified (#139, #96 and #23). The six features with negative average scores (relatively low risk), had scores from -0.87 to -0.56. The relationship between the DLS score of each feature with the 5-year AUC for the quantitation of each feature is presented in Supplementary Figure S6.

**Tumor-Adipose Feature**

The TAF finding was notable in several respects. First, across all clustering-derived features, TAF had the strongest association with high case-level and patch-level DLS scores. Second, case-level TAF quantitation (Supplementary Figure S7) was independently highly prognostic (Table 2, Figure 3B, Supplementary Table S4B, Supplementary Table S6B, Supplementary Table S7B). Given these results, we evaluated whether it was possible for researchers and pathologists to accurately identify TAF, thus enabling future work to better understand its biological and prognostic significance. Briefly, three non-anatomic-pathologists and two anatomic pathologists were presented with a total of 200 image patches from tumor-containing regions. For each patch, participants were instructed to indicate if that patch contained TAF or not. Accuracies for the non-pathologists were 90.0%, 93.0% and 95.5%, and accuracies for the pathologists were 87.0% and 90.5%. The inter-pathologist concordance was 93.5%.

# Discussion

In this study, we demonstrated the ability of a weakly-supervised deep learning system (DLS) to predict disease-specific survival (DSS) in intermediate stage colorectal cancer directly from unannotated, routine histopathology slides. We then developed a method for generating human-interpretable histologic features by clustering embeddings from a deep-learning based image-similarity model. We used these clustering-derived features, which explained a large fraction of the variance in DLS predictions, to gain an understanding of the histologic features the DLS scored as high and low risk. We found that one particular clustering-derived feature, characterized by poorly differentiated tumor cell clusters adjacent to adipose tissue, was strongly associated with high DLS risk scores, independently associated with poor prognosis and able to be reproducibly identified by pathologists.

We conducted a variety of statistical analyses that demonstrated the high prognostic performance of the DLS. First, the DLS provided significant risk stratification even within stage II and stage III cases. Furthermore, the difference in 5-year survival rates between high and low risk groups defined by the DLS was comparable to or greater than currently used prognostic factors such as obstruction, T-category, tumor-infiltrating lymphocytes, desmoplasia, lymphovascular invasion, and perineural invasion [11,27–31]. In multivariable analysis, the DLS added significant prognostic value to a set of 9 clinicopathologic baseline features. These results held across two validation datasets, including a temporal validation set from a later time period. These findings represent generalization of DLS performance, even to a cohort of cases with significant differences in baseline characteristics (Supplementary Table S1) as well as potential differences in treatment and technical aspects of tissue and slide preparation. Finally, the DLS performance was similar to that recently reported by Skrede et al.[4] using a comparable weakly supervised approach, further validating that substantial risk stratification is achievable with this type of deep learning approach.

Given the demonstrated ability of the DLS to risk-stratify patients, there is a potential for the DLS to inform clinical decisions involving the use of adjuvant chemotherapy. Specifically, the DLS could help identify high-risk stage II patients most likely to benefit from therapy or inform decisions about therapy regimens for low-risk stage III patients in order to minimize overtreatment. Prospective studies to evaluate the impact of DLS-informed treatment decisions on patient outcome are warranted, especially when combined with existing biomarkers that may provide complementary prognostic value.

Explainability is an important aspect of building the trust and transparency necessary for the adoption of such model-informed clinical decision making. This is especially true for weakly supervised prognostic models which learn to associate histologic features in unannotated whole-slide histopathology images without any human supervision. Although some insights have been derived from characterizing saliency heatmaps or example patches with extreme risk scores [21], researchers' ability to systematically characterize the histologic features learned by their model and evaluate the extent to which these features actually explain the model predictions remains limited.

While prior work has described weakly-supervised prognostic models for colorectal cancer with comparable performance to our DLS [4], an important advance offered by our study is the development of a computational method for generating human-interpretable "clustering-derived" features that can explain the DLS risk scores. We showed that while a set of 9 clinicopathologic features explained only a small fraction of variance in DLS scores (less than 20%, Table 4A), a set of 10 clustering-derived features, which could be understood, described, and reproducibly identified by pathologists, explained the majority of variance in DLS scores (about 60%, Table 4B). Finally, the complete set of 200 features explained another 15-20% of the variance in the DLS. This means approximately 20% of the variance remained unexplained, suggesting some features remained unappreciated by our method, and avenues for future work.

Although some of the features learned by weakly supervised prognostic models may be well known, there is also the possibility of learning previously unappreciated prognostic features. The clustering-derived feature most strongly associated with high DLS risk scores and poor prognosis was notable for its distinctive histomorphological appearance, including moderately to poorly differentiated tumor cells in close proximity to adipocytes, thus termed “Tumor Adipose Feature” (TAF). One initial interpretation might be that this feature represents invasion into the subserosa (T3 of TNM staging) or beyond (T4), and thus that the model may have learned a representation of the T-category, which has known prognostic significance [1]. However, both the DLS prediction and TAF quantitation remain significantly associated with survival even within T3 cases (Supplementary Table S6), suggesting prognostic value independent of T-category.

A hypothesis that could explain the independent prognostic value of TAF is submucosal adipose tissue as a prognostic factor itself, potentially associated with inflammatory bowel disease or obesity [32,33]. In regards to obesity, there is some evidence to suggest that body-mass index, visceral fat, and subcutaneous fat may be associated with adverse outcomes in metastatic colorectal cancer [34]. More speculatively, this finding may be consistent with an adverse role for cancer-associated adipocytes in colorectal cancer, as has been described in other cancer types [35,36]. Finally, there are notable morphologic similarities between TAF and irregular tumor growth at the invasive edge, potentially representing an association with “infiltrative” versus “pushing” configurations of the tumor border [37,38]. Further work is warranted to better understand the biological significance of TAF and other clustering-derived features.

Our study has some limitations. First, as a retrospective study, treatment pathways present an important confounding factor that is difficult to control for, including potential differences in neoadjuvant and adjuvant therapy. Though treatment guidelines within stage II and within stage III colorectal cancer cohorts are fairly uniform, at least some variability in treatment likely exists. Progression-free survival may be an endpoint that is less susceptible to treatment confounding, but was unfortunately not available at the scale required for this study. Second, while the non-random temporal validation set demonstrates generalization in the face of significant changes in case characteristics over time (Supplementary Table S1), validation in geographically diverse cohorts would be needed to further support the generalization of the DLS to other cohorts containing complete, routine clinical cases. Unfortunately, such geographically diverse data with the necessary imaging and clinical data were not available for this study. A further limitation is that we were not able to evaluate the association between the DLS and several known prognosis factors such as tumor budding, number of lymph nodes examined, tumor location, obstruction, microsatellite instability, tumor-infiltrating lymphocytes, molecular profile (eg, BRAF and KRAS), desmoplasia, or histologic subtypes [11,30,31,39,40]. While obvious associations with TILs, desmoplasia, or subtype were not observed in our analysis of clustering-derived features, the association of the DLS scores with these factors will need to be examined in future work. Though used in our analysis, lymphovascular invasion was not formally re-evaluated for the purposes of this study and thus may not be exhaustively recorded. While we were able to show that individual patches containing TAF can be reproducibly identified, suggesting that the feature is readily learnable, further work is required to validate the prognostic value of pathologists’ case-level quantitation of TAF. Doing so will require the development of guidelines to ensure consistent scoring across pathologists. While the use of a clustering algorithm facilitated the identification of TAF, the clusters themselves are based on image similarity rather than specific histopathological concepts. Thus, in building on the methods and findings here, pathologist-guided refinement of algorithm-derived feature clusters may lead to even more prognostic and well-defined features. Finally, the cluster analysis provided valuable insights into the features that could explain the variance in DLS scores, but there may be additional important features that were not identified by these specific clusters. For example, generating clusters using embeddings from different machine learning models [25] could potentially help identify additional features that further explain DLS predictions.

To conclude, the present work demonstrates the application of deep learning methods to learn and describe histomorphologic features with prognostic value for colorectal cancer, without pre-specification of features. The prognostic predictions of the DLS provided significant risk stratification in both stage II and stage III cases, even after adjusting for a number of clinicopathologic features including T category, N category, and tumor grade. Individual histologic features associated with risk predictions by the DLS were also characterized, providing a framework for future efforts in explaining weakly supervised models in histopathology. Finally, this analysis enabled the description and reproducible identification of a visually-distinctive machine-learned feature with independent prognostic significance. This ability to learn from machine learning represents an important first step in allowing experts to further study new concepts discovered using weakly supervised deep learning models.

# Methods

**Data Cohorts**

This study utilized archived formalin-fixed paraffin-embedded, hematoxylin and eosin stained pathology slides from the Institute of Pathology and the BioBank at the Medical University of Graz [41]. Institutional Review Board approval for this retrospective study using de-identified slides was obtained from the Medical University of Graz [42,43] (Protocol no. 30-184 ex 17/18). All available slides in archived stage II and stage III colorectal cancer resection cases between 1984 and 2013 were retrieved, de-identified, and scanned using a Leica Aperio AT2 scanner at 20X magnification (0.5 µm/pixel). The complete set of digitized whole slide images (WSIs) consisted of 6,437 cases and 114,561 slides. Additional de-identified clinical and pathological information corresponding to each case were extracted from pathology reports [44,45] along with data from Statistik Austria. This information included pathologic TNM staging, tumor grade (G), resection margin status (R), sex, and age at diagnosis. When indicated in the report, presence of lymphatic invasion (L) and venous invasion (V) were also extracted. Disease-specific survival (DSS) was inferred from the International Classification of Diseases (ICD) code available for cause of death and only colorectal cancer associated ICD codes were considered disease-specific (C18, C19, C20, C21, C26, C97), with other types of cancer excluded.

All 114,561 slides underwent manual review by pathologists to identify the stain and tissue type. Immunohistochemistry-stained slides and non-colorectal specimens such as lymph node, small intestine, and other tissue types, were excluded. In addition, cases with low tumor content, death within 30 days of surgical resection and secondary tumor resections were also excluded, leaving 43,780 slides from 5,629 cases (Supplementary Figure S1). These slides were partitioned into two cohorts. All cases from 1984-2007 were assigned to the first cohort, which was randomly subdivided into a training set, a tuning set, and the first validation set in a 2:1:1 ratio. To further evaluate the performance of the model and assess temporal generalizability, all cases from 2008-2013 were assigned to the second validation set. This division of years was used to ensure 5 years of followup were available for all cases, and that validation set 2 contained an arbitrarily-determined 5 years worth of cases. Validation set 1 contains 1,239 cases with 9,340 slides while validation set 2 contains 738 cases with 7,140 slides (Table 1). The distributions of clinical metadata in the validation sets are described in Supplementary Table S1, and the Kaplan-Meier curves for all splits are shown in Supplementary Figure S8.

**Deep Learning System Overview**

The DLS consisted of two separate models applied sequentially. First, a tumor segmentation model was applied on each whole-slide image (WSI) to generate a region of interest (ROI) mask. A prognostic model was then trained and evaluated to predict case-level DSS using image patches sampled from these ROI masks (Supplementary Figure S2).

**Tumor Segmentation Model**

In order to identify tumor-containing regions at scale, we first developed a model for colorectal adenocarcinoma detection using an approach similar to that previously described [46]. Briefly, a convolutional neural network (CNN) was developed in a patch-based supervised learning approach using WSIs from 200 pathologist-annotated colorectal slides. These annotations involved pixel-level outlines of colorectal adenocarcinoma, normal colorectal epithelium, atypical epithelium, and non-epithelial tissue. The model was developed solely using our development dataset and achieved an AUC of 98.50. This colorectal adenocarcinoma detection model was used to generate ROI masks. Only patches from within the ROI masks were used to train and evaluate the prognostic model. Additional details on the development of the tumor segmentation model and the generation of ROI masks are in the Supplementary Methods and Supplementary Tables S8 and S9.

**Prognostic Model Neural Network Architecture and Survival Loss**

The neural network architecture for the prognostic model was designed to predict a case-level risk score given a set of image patches sampled from the tumor containing regions in a case, and was previously described [47]. The architecture consisted of several CNN modules with shared weights for extracting dense feature vectors from each input patch, an average pooling layer for merging the set of patch-level feature vectors into a single case-level feature vector, and a final Cox regression layer for computing a scalar case-level risk score (see Supplementary Figure S2C). The CNNs consisted of depth-wise separable convolution layers, similar to the design of MobileNet [47,48]. This type of convolution layer has fewer parameters than standard convolution layers, which reduces computation and helps avoid overfitting. The filter size in each layer and the number of layers were tuned via a random grid-search [49] (Supplementary Table S10).

For the weakly-supervised survival prediction tasks, the location of informative patches in each WSI is not known. Our approach of randomly sampling $n$ patches within each slide helped ensure informative patches were selected during training. If each patch has a certain probability of being informative, the probability of not sampling any informative patches decreases exponentially with the increase of $n$. This approach also generalizes to different numbers of slides per case, enabling use in real world datasets that may contain many slides per case (average of 18 slides/case in our study).

The loss function during training was the Cox partial likelihood [50], which was selected based on a preliminary experiment on the tune set where it performed the best (by a small margin) amongst the three survival loss functions (Supplementary Figure S9). By contrast, in our prior work with different cohorts, different inclusion criteria, and only one slide per case[47], the censored cross-entropy loss function performed better, indicating value in further work to better understand the optimal loss function. The Cox partial likelihood is formulated as follows:

$$\max \prod_{i:O_i=1} \frac{e^{f(X_i)}}{\sum_{j:T_j \geq T_i} e^{f(X_j)}}$$ --- Equation (1)

where for the $i^{th}$ case, $T_i$ is the event time or time of last follow-up, $O_i$ is an indicator variable for whether the event is observed, $X_i$ is the set of WSIs. The function $f$ represents the prognostic model, and $f(X_i)$ is the scalar case-level risk score. In our implementation, we used Breslow's approximation [51] for handling tied event times due to its simplicity of implementation. During training, we approximated the full loss at each training step by evaluating it over the examples in the training batch.

**Prognostic Model Training Procedure**
The prognostic model was trained on both stage II and stage III cases. Training examples consisted of sets of 16 image patches per case sampled randomly across regions of interest produced by the ROI model. Images were first normalized to a standard color distribution based on the color statistics in the training set [46] and then augmented by color and orientation perturbations described previously [46]. Numerical optimization of network parameters was done using the Adam optimizer [52]. Hyperparameters governing ROI mask generation, patch extraction, model architecture and optimizer were tuned by selecting the best performing configuration across 100 random configurations from the full hyperparameter search space (Supplementary Table S10). Models were trained for 2 million steps in a distributed fashion, using 50 workers with 16 CPU processors each.

**Prognostic Model Evaluation Procedure**
Each model was evaluated every 10,000 steps on the tuning set using a sample of 1,024 patches per case. The best checkpoint for each model was selected by taking the maximum after applying a rolling average with a window size of 10. The best checkpoints for five models that achieved the highest c-index on the tuning set were ensembled to form the final prognostic model. To generate a case-level prognostic risk score, the ensembled prognostic model was run exhaustively over all non-overlapping patches within the ROI mask.

**Evaluating DLS Performance**
We used three evaluation metrics to assess the prognostic ability of the DLS for DSS: 5-year survival AUC, hazard ratio, and c-index. These analyses were pre-specified and documented prior to running the model on the validation sets. All analyses were done for stage II and stage III independently, and for the two stages combined. The 5-year AUC was used because every case in both validation sets had at least 5 years of followup. In the two validation sets, 10% of examples were censored prior to 5 years due to non-disease-specific death; these examples were excluded for the purposes of 5-year DSS AUC computation, but incorporated as right-censored for hazards ratio and c-index computation. The 5-year AUC for the clinicopathological variables and the combination of these clinicopathological variables with the DLS was computed using the sklearn.metrics.roc_auc_score function in the Python sklearn package (v0.23.2).

To compute the hazard ratio for the DLS as well as the clinicopathological variables, Cox proportional hazards regression models [53] were used. The case-level DLS scores and age were treated as numeric variables. DLS predictions were rescaled to have zero mean and unit variance. Age was centered at the mean age scaled down by a factor of 10, such that the hazard ratio for age corresponds to the risk increase per decade of age. All other variables were coded as categorical (dummy/indicator) variables. Survival times were discretized into months for all analyses.

Cox regression models were also used to calculate c-indices [54] for the DLS, the clinicopathologic features alone (baseline model), and for the DLS combined with these variables (combined model). These Cox models were fit on the tune set and applied to both validation sets. C-indices were computed using the lifelines.utils.concordance_index() function in the Python Lifelines package (v0.24.6). Confidence intervals for the c-index were generated via paired bootstrap resampling with 9,999 samples.

For Kaplan-Meier analysis, cases were stratified into low and high risk groups using thresholds determined on the tune set. The low risk threshold is the 25th percentile of tune set risk scores, while the high risk threshold is the 75th percentile of tune set risk scores. Different thresholds were selected for stage II cases, stage III cases and the combination of stage II and stage III cases. To account for the temporal shift in case characteristics shown in Supplementary Table S1, only tune set cases from the most recent 5 years (2002-2007) were used for selecting the threshold for validation set 2 (all cases from 2008-2013). The Python Lifelines package (version 0.24.6) [55] was used for Kaplan Meier analysis and Cox regression analyses, using the lifelines.KaplanMeierFitter and lifelines.CoxPHFitter classes. The REMARK checklist for reporting is provided as Supplementary Table S11.

**Understanding DLS Predictions**

The following analyses were conducted in an exploratory manner after the DLS was applied to the validation sets. All annotations and histologic reviews were performed with the raters blinded to both the DLS's predictions and the outcomes of the relevant case.

**DLS Association with Clinicopathologic Features**

The association of the DLS with case-level clinicopathologic features was evaluated via multivariable linear regression (Table 4A). The case-level DLS scores were standardized to have zero mean and unit variance. All clinicopathologic features except age (at the time of diagnosis) were coded as indicator variables. Age was centered at the mean age and scaled down by a factor of 10, such that the coefficient for age corresponds to the risk increase per decade of age. The proportion of variance in DLS scores explained by these features was evaluated using the adjusted coefficient of determination ($R^2$).

**DLS Association with Clustering-Derived Features**

We next studied the association of the DLS with histologic features derived from clustering (Table 4B). To obtain these histologic features, we leveraged a previously-described image-similarity deep learning model [25,26] that was trained to distinguish between similar and non-similar natural (non-histopathology) images. This model was used to generate patch-level embeddings that captured visual similarity. Embeddings for a sample of 100,000 tumor-containing training set image patches were clustered using the k-means algorithm as implemented in the Python sklearn package (v0.21.3).

The total number of clusters (k) was chosen based on the fit on the tune set (described next) when using the best subset of 10 features (described next). Values of k explored were: 10, 25, 50, 100, 200, 300, 400, and 500. K=200 clusters was found to be optimal. The centroids of these 200 clusters, which were fit on the sample of patches from the training set, were used to assign each tumor-containing patch in both validation sets to a cluster. For each case, the percentage of patches belonging to each cluster was computed.

Next, the association of the DLS with these features was also evaluated via multivariable linear regression in a similar manner to the procedure for clinicopathologic features above. The clustering-derived features were scaled to range from 0 to 100, indicating the percentage of tumor in the case belonging to each feature. A subset of 10 features was selected for more in-depth characterization from the full set of 200 features using forward stepwise selection with the objective of maximizing the $R^2$.

To provide morphological descriptions for the subset of 10 features, 15 patches per feature were presented independently to two pathologists for review (a subset of 10 per feature are shown in Fig. 2). The selected patches were those that were closest to each feature's centroid (and filtered to ensure that for each feature, each patch was sampled from a different case). The pathologists were blinded to any additional information about the feature, and provided histopathological review via a structured form. The presence of tumor, stroma, adipose, and TILs were scored semi-quantitatively as absent, low, medium, or high. Tumor was graded as low, intermediate, or high grade, and fibrosis (if present) was graded as mature, intermediate, or immature [56]. Additional free-text descriptions of tumor and stroma for each cluster were also provided by each pathologist.

**DLS Association with Patch-Level Histoprognostic Features**

To evaluate the association of DLS predictions with known histoprognostic features, we annotated 161 slides for several known features previously reported to be associated with adverse prognosis in colorectal cancer. The slides used for this purpose were randomly selected from 161 cases in validation set 2. Annotated features included lymphovascular invasion, perineural invasion, intratumoral budding, peritumoral budding, and peritumoral fibrosis. When present, polyps were also annotated to provide another histologic class for comparison. Board-certified pathologists (without gastrointestinal subspecialty training, median pathologist experience: 6.5 years post-training, range 3-17 years) were asked to exhaustively annotate the tumor-containing regions of each slide for these features.

For each histoprognostic feature, the average patch-level DLS score among all patches annotated for that feature was computed. For comparison, we also computed the average patch-level DLS score among all patches for each clustering-derived feature. For both analyses 95% confidence intervals for the average DLS score were computed via blocked bootstrapping at the slide-level.

**Tumor-Adipose Feature**

To understand if people could accurately identify TAF, we extracted both TAF-containing and non-TAF-containing image patches from tumor-containing regions (based on the tumor segmentation model). Each patch was 256 × 256 pixels at 5X magnification (0.5 $mm^2$). Participants were first presented 50 TAF patches (as determined by the clustering algorithm) as learning material. Of these, 25 (Figure 3A) were closest to the centroid and thus the most representative of the cluster-derived feature. Another 25 patches were randomly sampled from the cluster. These randomly sampled patches potentially included examples without the pathologist-identified tumor adipose feature, and were included to provide examples of the diversity of patches assigned to the cluster. Within each set of 25, each patch came from a distinct case. The participants (two anatomic pathologists: I.F-A. and T.B. and three non-anatomic-pathologists: E.W., D.F.S., and Y.L.) reviewed the above material, and then completed a separate practice round of indicating whether they perceived each of 50 additional patches to be TAF or not. Clustering algorithm labels were subsequently provided as feedback. Finally, we prepared an independent set of 200 patches, of which 100 were randomly sampled from all patches classified by the clustering algorithm as TAF, while the remaining 100 were randomly sampled from all patches not classified as TAF. The participants again indicated whether each patch was TAF or not. All patches and their labels (TAF vs not based on the clustering algorithm) are provided as supplementary material. To avoid biasing this study of the cluster-derived feature, these patches were not otherwise filtered or reviewed by a pathologist to fit any annotator's mental concept of TAF. As an additional exploratory analysis, we also generated TAF patches by finding cluster centroids using validation set 2 instead of validation set 1, with similar results (Supplementary Figure S10).

**Model Inference Speed**
The inference timings per case are: 11±7 minutes (±standard deviation) for a single machine with 16 cores; 13±8 seconds for 50 such machines in a cloud environment; and 8±5 seconds for a commercially-available accelerator, Google Cloud Tensor Processing Unit (v2). These timings range from being comparable to significantly faster than slide preparation and digitization, which can take a few minutes per slide (multiplied by about 10 slides per case on average; See Table 1).

# Data availability

This study utilized archived anonymized pathology slides, clinicopathologic variables, and outcomes from the Institute of Pathology and the BioBank at the Medical University of Graz. Interested researchers should contact K. Z. to inquire about access to Biobank Graz data; requests for non-commercial academic use will be considered and require ethics review prior to access.

# Acknowledgements

This work was funded by Google LLC and Verily Life Sciences. The authors would like to acknowledge the Pathology team and in particular Timo Kohlberger, Yuannan Cai, Hongwu (Harry) Wang and Angela Lin for software infrastructure support and data collection. We also appreciate the input of Kunal Nagpal, Akinori Mitani, and Dale Webster for their feedback on the manuscript. Last but not least, this work would not have been possible without the support of Dr. Christian Guelly, the Biobank Graz, the efforts of the slide digitization team at the Medical University Graz and the participation of the pathologists who reviewed the cases for quality control or to annotate tumor and known prognostic features.

# Competing interests

E.W., D.F.S., M.M., F.T., P-H.C.C., N.H., A.S., R.M., B.A., G.S.C., L.H.P., D.T., Z.X., Y.L., M.C.S., and C.H.M. are current or past employees of Google LLC and own Alphabet stock. I.F-A. and T.B. are consultants of Google LLC. M.P., R.R., P.R., H.M., and K.Z are employees of the Medical University of Graz.

# Author contributions

E.W. performed the majority of the machine learning development and validation with input from D.F.S, P-H.C.C., Z.X., Y.L., and M.C.S.. E.W., A.S., and Z.X. wrote the technical infrastructure needed for machine learning; D.F.S., M.M., M.P., R.R., F.T., I.F-A,., T.B., R.M., and B.A. collected and performed quality control for the data; I.F-A., T.B., P.R., and K.Z. interpreted the clusters; E.W., D.F.S., and N.H. prepared data for the known prognostic feature analysis. G.S.C., L.H.P., D.T., H.M., M.C.S., K.Z., and C.H.M. obtained funding for data collection and analysis, supervised the study, and provided strategic guidance. E.W., D.F.S., and Y.L. prepared the manuscript with input from all authors. K.Z. and C.H.M. contributed equally.

# Code availability

The deep learning framework (TensorFlow) used in this study is available at https://www.tensorflow.org/; the deep learning architecture is provided as detailed pseudocode in S1 algorithm at https://journals.plos.org/plosone/article?id=10.1371/journal.pone.0233678 and leverages standard depth-wise separable convolutions available from Keras as SeparableConv2D: https://github.com/tensorflow/tensorflow/blob/v2.1.0/tensorflow/python/keras/layers/convolutional.py.

# Figures

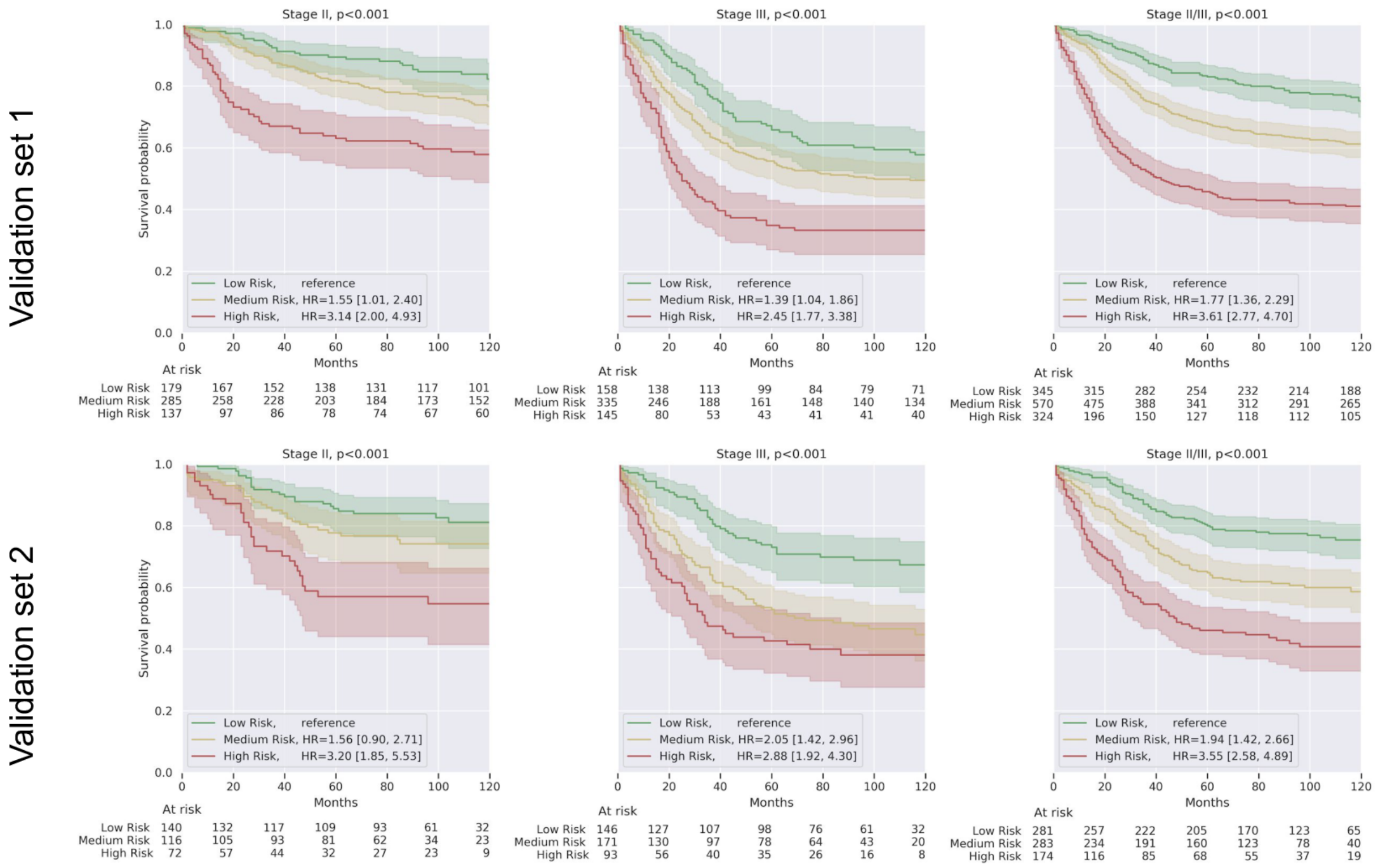

**Figure 1. Kaplan Meier curves on both validation sets for patients stratified by the prognostic deep learning system (DLS).** Results are presented for stage II and stage III patients separately, and as a combined cohort (Stage II/III). High and low risk groups represent the highest and lowest risk quartiles, respectively, based on the DLS prediction. Hazard ratios (HR) for the medium and high risk groups are provided with the low-risk group as the reference group. Shaded areas represent 95% confidence intervals. P-values were calculated using the log-rank test comparing each high risk group with the corresponding low risk group.

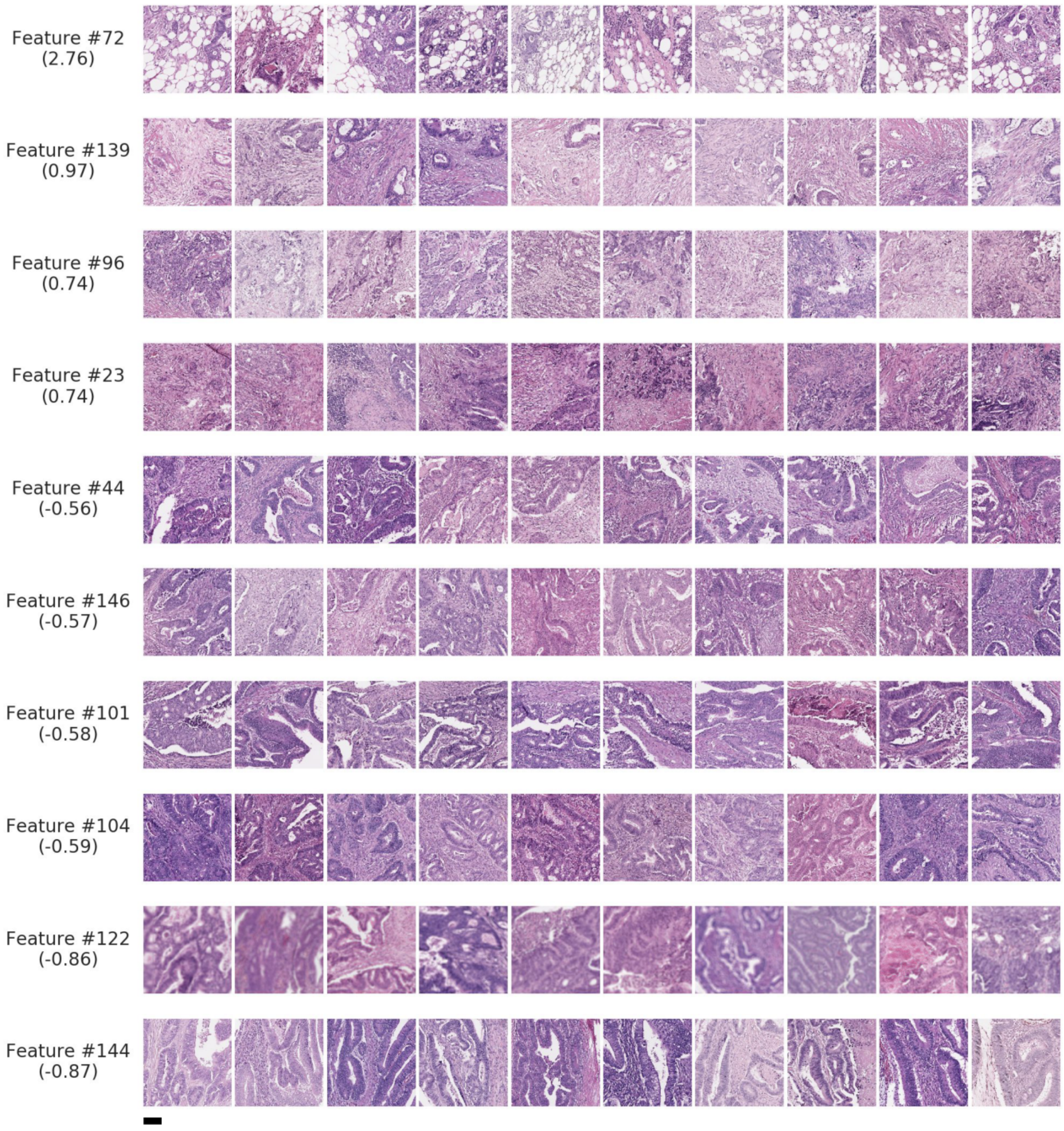

**Figure 2. Representative patches for clustering-derived features associated with predictions of the deep learning system (DLS).** Sample patches for a set of 10 clustering-derived features are shown. For each feature, the 10 patches closest to the centroid were selected, after filtering to ensure they were from distinct cases (Methods). The case-level quantitation of these 4 high-risk and 6 low-risk features explains the majority of the variance in case-level DLS scores. Features are ranked according to the average DLS score, which is provided in parentheses. Scale bar indicates 0.1 mm.

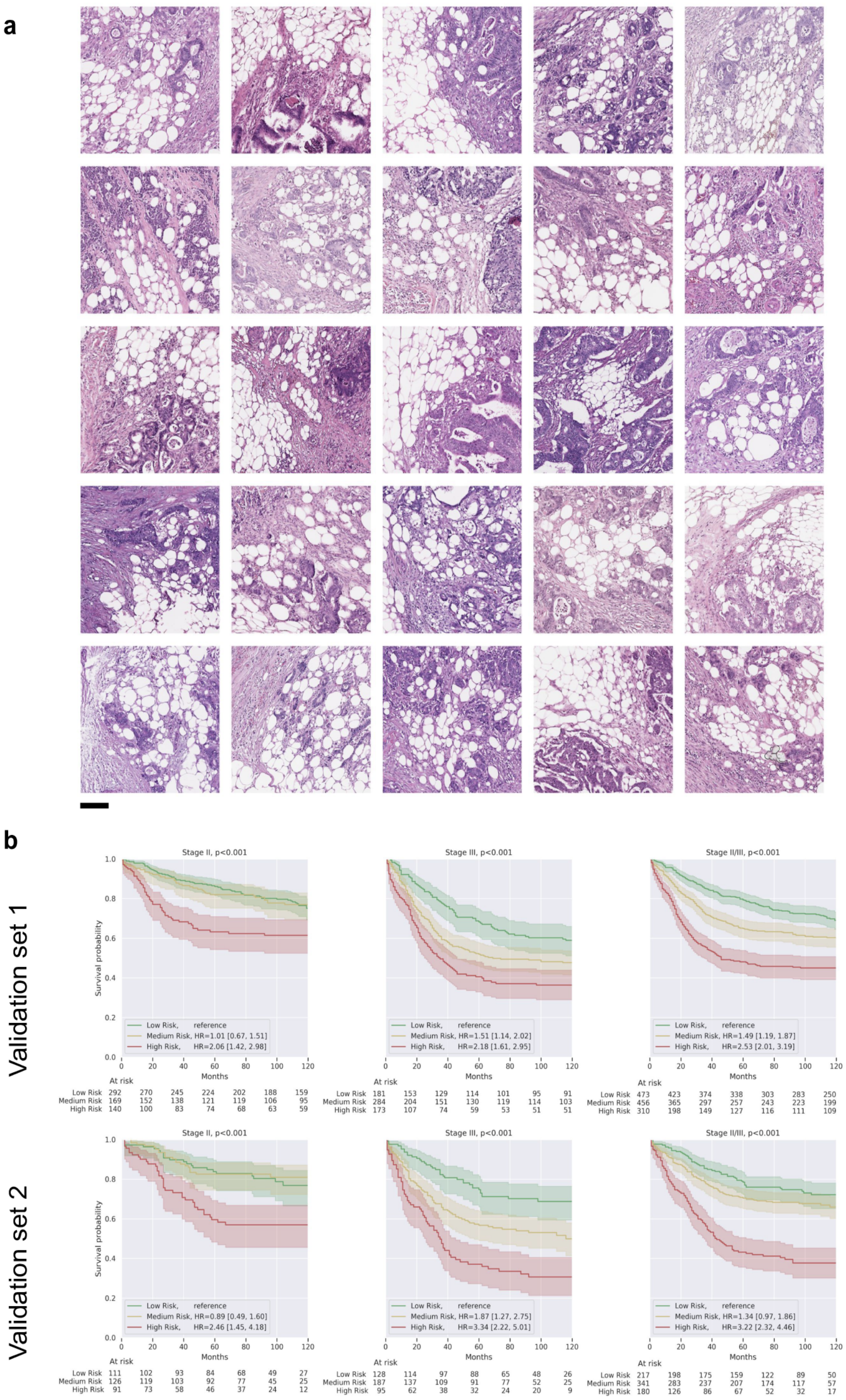

**Figure 3. Visualizations and survival analysis of the clustering-derived feature with the highest DLS-predicted risk score (tumor-adipose feature, TAF).** (a) Additional sample patches of the TAF cluster, each from a unique case. Scale bar indicates 0.1 mm. (b) Kaplan Meier curves on both validation sets for patients stratified by quantitation of TAF. These curves were generated following the same procedure as in Figure 1. In stage II cases, the deviation in at-risk counts from the quartile marks for the low-risk and medium-risk groups are because many stage II cases (50% in validation set 1 and 38% in validation set 2) did not contain any TAF.

# Tables

**Table 1. Data used in this study**

All cases were from the Institute of Pathology and the BioBank at the Medical University of Graz. Cases between 1984-2007 were randomly split in a ratio of 3:1 into a development set and validation set 1. The development set was further split into train and tune sets in a 2:1 ratio. Additional cases from 2008-2013 were obtained after model development as validation set 2. Disease-specific survival (DSS) was inferred from the International Classification of Diseases (ICD) code available for cause of death. Only slides containing colorectal tissue were used for development and validation.

| **Study** | **No. of cases** | | | | **No. of DSS events (%)** | | | | **No. of slides** | | | |
|---|---|---|---|---|---|---|---|---|---|---|---|---|
| | Train | Tune | Validation set 1 | Validation set 2 | Train | Tune | Validation set 1 | Validation set 2 | Train | Tune | Validation set 1 | Validation set 2 |
| **Stage II** | 1173 | 586 | 601 | 328 | 303 (26%) | 152 (26%) | 152 (25%) | 80 (24%) | 8687 | 4205 | 4452 | 3227 |
| **Stage III** | 1266 | 627 | 638 | 410 | 609 (48%) | 294 (47%) | 312 (49%) | 183 (45%) | 9617 | 4791 | 4888 | 3913 |
| **Stage II/III** | 2439 | 1213 | 1239 | 738 | 912 (37%) | 446 (37%) | 464 (37%) | 263 (36%) | 18304 | 8996 | 9340 | 7140 |

**Table 2. The 5-year AUC for disease-specific survival (DSS) prediction**

| Cancer stage | Dataset | DLS | Quantitation of tumor-adipose feature |
|---|---|---|---|
| Stage II | Validation set 1 (n=601 cases) | 0.680 [0.631, 0.739] | 0.645 [0.598, 0.700] |
| | Validation set 2 (n=328 cases) | 0.663 [0.592, 0.730] | 0.634 [0.570, 0.697] |
| Stage III | Validation set 1 (n=638 cases) | 0.655 [0.617, 0.694] | 0.629 [0.593, 0.680] |
| | Validation set 2 (n=410 cases) | 0.655 [0.600, 0.707] | 0.682 [0.638, 0.743] |
| Stage II/III | Validation set 1 (n=1,239 cases) | 0.698 [0.660, 0.729] | 0.661 [0.629, 0.694] |
| | Validation set 2 (n=738 cases) | 0.686 [0.638, 0.723] | 0.682 [0.641, 0.734] |

## Table 3. Multivariable Cox regression on the validation sets

Numbers indicate hazard ratio followed by 95% confidence intervals in square brackets, and p-values (from a Wald test) after the comma. Corresponding univariable analysis is presented in Supplementary Table S3. *N/A because stage II only contains N0 and T3 or T4 and stage III only contains N1 by definition (American Joint Committee on Cancer, AJCC). Bold indicates statistically significant input variables (p < 0.05).

| Variable | Stage II | | Stage III | | Stage II/III | |
|---|---|---|---|---|---|---|
| | Validation set 1 | Validation set 2 | Validation set 1 | Validation set 2 | Validation set 1 | Validation set 2 |
| **DLS** | **1.64 [1.39, 1.93], <0.001** | **1.54 [1.22, 1.94], <0.001** | **1.42 [1.26, 1.61], <0.001** | **1.39 [1.20, 1.61], <0.001** | **1.54 [1.38, 1.70], <0.001** | **1.42 [1.25, 1.61], <0.001** |
| **Age** | 1.13 [0.96, 1.33], 0.128 | **1.49 [1.17, 1.89], <0.001** | **1.15 [1.05, 1.26], 0.004** | **1.25 [1.10, 1.43], <0.001** | **1.14 [1.05, 1.24], 0.002** | **1.31 [1.17, 1.47], <0.001** |
| **Sex** | | | | | | |
| **Male** | 1.0 (reference) | | | | | |
| **Female** | **0.69 [0.49, 0.96], 0.028** | 0.75 [0.47, 1.21], 0.240 | **0.76 [0.60, 0.95], 0.017** | 0.96 [0.71, 1.29], 0.766 | **0.74 [0.61, 0.89], 0.002** | 0.89 [0.70, 1.14], 0.360 |
| **Grade** | | | | | | |
| **G1** | 1.0 (reference) | | | | | |
| **G2** | 0.80 [0.39, 1.66], 0.550 | 1.38 [0.49, 3.88], 0.536 | 1.16 [0.51, 2.64], 0.719 | 2.80 [0.87, 8.99], 0.083 | 0.96 [0.56, 1.66], 0.897 | 1.98 [0.92, 4.25], 0.082 |
| **G3** | 0.88 [0.40, 1.96], 0.756 | 0.99 [0.33, 3.00], 0.990 | 1.47 [0.63, 3.39], 0.372 | 2.88 [0.89, 9.28], 0.077 | 1.19 [0.68, 2.08], 0.550 | 1.86 [0.85, 4.07], 0.119 |
| **GX** | 0.92 [0.19, 4.36], 0.916 | 1.37 [0.25, 7.61], 0.718 | 0.73 [0.15, 3.68], 0.707 | 2.56 [0.60, 10.93], 0.204 | 0.78 [0.25, 2.37], 0.657 | 1.86 [0.65, 5.36], 0.250 |
| **Lymphatic Invasion** | | | | | | |
| **L0** | 1.0 (reference) | | | | | |
| **L1** | 1.40 [0.88, 2.24], 0.154 | 0.74 [0.39, 1.41], 0.352 | 0.80 [0.60, 1.08], 0.146 | 0.99 [0.71, 1.38], 0.948 | 0.95 [0.74, 1.22], 0.692 | 0.92 [0.69, 1.23], 0.568 |
| **N-category** | | | | | | |
| **N0** | N/A* | | | | 1.0 (reference) | |
| **N1** | N/A* | | 1.0 (reference) | | **1.89 [1.49, 2.39], <0.001** | **1.68 [1.23, 2.29], 0.001** |
| **N2** | N/A* | | 1.10 [0.85, 1.42], 0.482 | 1.29 [0.95, 1.76], 0.107 | **2.03 [1.55, 2.67], <0.001** | **2.21 [1.58, 3.08], <0.001** |
| **N3** | N/A* | | 1.03 [0.73, 1.47], 0.858 | 0.60 [0.15, 2.46], 0.481 | **1.85 [1.29, 2.66], 0.001** | 1.02 [0.25, 4.18], 0.973 |
| **Margin Status** | | | | | | |
| **R0** | 1.0 (reference) | | | | | |
| **R1** | 1.22 [0.44, 3.38], 0.700 | 1.01 [0.30, 3.39], 0.982 | 1.08 [0.66, 1.77], 0.761 | 0.74 [0.36, 1.54], 0.419 | 1.10 [0.71, 1.72], 0.666 | 0.81 [0.44, 1.50], 0.503 |
| **T-category** | | | | | | |
| **T1/T2** | N/A* | | 1.0 (reference) | | | |
| **T3** | 1.0 (reference) | | 1.37 [0.88, 2.12], 0.159 | **2.36 [1.09, 5.09], 0.029** | 1.29 [0.84, 2.00], 0.244 | **2.31 [1.07, 4.98], 0.032** |
| **T4** | 1.53 [0.95, 2.47], 0.081 | **1.86 [1.08, 3.22], 0.026** | **1.66 [1.02, 2.71], 0.042** | **4.36 [1.98, 9.59], <0.001** | **1.66 [1.03, 2.65], 0.037** | **4.25 [1.95, 9.28], <0.001** |
| **Venous Invasion** | | | | | | |
| **V0** | 1.0 (reference) | | | | | |
| **V1** | 1.37 [0.67, 2.80], 0.389 | 1.42 [0.69, 2.95], 0.345 | 0.74 [0.48, 1.13], 0.165 | 1.18 [0.82, 1.69], 0.369 | 0.82 [0.57, 1.18], 0.278 | 1.20 [0.87, 1.65], 0.270 |

**Table 4. Multivariable regression of case-level DLS score using (a) clinicopathologic features and (b) clustering-derived features as input**

P-values (from a t-test) for the overall model in both panels A and B are <0.001. Each coefficient represents the relative increase of the DLS score associated with that variable. Bold indicates statistically significant input variables ($p < 0.05$).

**a**

| Clinicopathologic feature | Validation Set 1 | | | Validation Set 2 | | |
|---|---|---|---|---|---|---|
| | Coefficient | p | $R^2$ | Coefficient | p | $R^2$ |
| T3 | **0.5454** | **<0.001** | 0.18 | 0.1184 | 0.276 | 0.18 |
| T4 | **0.7775** | **<0.001** | | **0.4032** | **<0.001** | |
| N1 | **0.5496** | **<0.001** | | **0.2912** | **<0.001** | |
| N2 | **0.5942** | **<0.001** | | **0.4752** | **<0.001** | |
| N3 | **1.0311** | **<0.001** | | 0.3477 | 0.163 | |
| R1 | 0.1108 | 0.427 | | **0.3365** | **0.011** | |
| L1 | -0.1569 | 0.032 | | 0.1063 | 0.074 | |
| V1 | 0.2376 | 0.033 | | 0.1332 | 0.054 | |
| Grade 2 | 0.1032 | 0.467 | | 0.0557 | 0.605 | |
| Grade 3 | **0.4342** | **0.004** | | 0.1800 | 0.112 | |
| Grade X | 0.5504 | 0.049 | | 0.1968 | 0.287 | |
| Sex (female) | -0.0091 | 0.862 | | 0.0179 | 0.713 | |
| Age at diagnosis | **-0.0670** | **0.002** | | -0.0043 | 0.833 | |
| Intercept | **-1.0471** | **<0.001** | | **-1.4258** | **<0.001** | |

**b**

| Feature # | Description | Validation Set 1 | | | Validation Set 2 | | |
|---|---|---|---|---|---|---|---|
| | | Coefficient | p | $R^2$ | Coefficient | p | $R^2$ |
| 72 | Small clusters of moderate to high grade tumor cells intermixed with substantial adipose and a minor component of desmoplastic stroma | **0.2269** | **<0.001** | 0.57 | **0.2913** | **<0.001** | 0.61 |
| 139 | Low-intermediate grade tumor with predominant stroma of mature and intermediate desmoplasia | **0.1977** | **<0.001** | | **0.1650** | **<0.001** | |
| 23 | Small clusters of high grade tumor cells with predominant, mature desmoplasia and moderate TILs | **0.1096** | **<0.001** | | **0.1931** | **<0.001** | |
| 96 | Small clusters of high grade tumor cells, including single tumor cells, and moderate amount of mature and intermediate desmoplasia | **0.1031** | **<0.001** | | **0.1996** | **<0.001** | |
| 146 | Low grade tumor with moderate differentiation and desmoplastic stroma with mature desmoplasia and occasional TILs | **-0.1248** | **<0.001** | | **-0.2133** | **<0.001** | |
| 122 | Out of focus regions; predominantly low grade tumor with tubule formation. | **-0.1323** | **<0.001** | | -0.2867 | 0.187 | |
| 104 | Low and Intermediate grade tumor with tubule formation and small, solid regions; Stroma with mature desmoplasia | **-0.1461** | **<0.001** | | **-0.0505** | **<0.001** | |
| 44 | Intermediate grade tumor with irregular tubule formation; mature desmoplasia and focal areas of TILs | **-0.1510** | **<0.001** | | **-0.1081** | **<0.001** | |
| 101 | Predominantly Intermediate grade tumor with irregular tubule formation; minor component of mature, desmoplasia | **-0.2312** | **<0.001** | | -0.0420 | 0.313 | |
| 144 | Low grade tumor with tubule formation and minor component of mixed stroma containing mature and intermediate desmoplasia with occasional, moderate TILs | **-0.3476** | **<0.001** | | **-0.3794** | **<0.001** | |
| Intercept | N/A | **0.1256** | **0.002** | | **0.1996** | **<0.001** | |
| Adding remaining 190 features | N/A | N/A | N/A | 0.73 | N/A | N/A | 0.80 |

**Table 5. Average and interquartile range of DLS scores across patches for clustering-derived features and known histologic features**
Confidence intervals were computed via block bootstrapping.

| Source of feature | Feature name | DLS score mean (95% CI) | DLS score interquartile range |
|---|---|---|---|
| Known features (manually annotated by pathologists; 87,325 patches across 161 slides) | Lymphovascular invasion | 1.03 [0.33, 1.95] | [0.09, 1.82] |
| | Perineural invasion | 0.75 [0.14, 1.28] | [-0.18, 1.68] |
| | Intratumoral budding | 0.33 [0.00, 0.59] | [-0.63, 1.15] |
| | Peritumoral fibrosis | 0.26 [0.02, 0.42] | [-0.73, 1.18] |
| | Peritumoral budding | 0.10 [-0.10, 0.30] | [-0.96, 0.94] |
| | Other adenocarcinoma | -0.46 [-0.57, -0.36] | [-1.36, 0.25] |
| | Polyp | -0.86 [-1.26, -0.59] | [-1.57, -0.24] |
| Clustering-derived (from clusters identified by a deep learning-based visual similarity model; 2,568,691 patches across 9,340 slides) | 72 | 2.76 [2.59, 2.93] | [1.66, 3.74] |
| | 139 | 0.97 [0.91, 1.02] | [0.40, 1.61] |
| | 96 | 0.74 [0.69, 0.80] | [0.13, 1.42] |
| | 23 | 0.74 [0.68, 0.78] | [0.08, 1.38] |
| | 44 | -0.56 [-0.61, -0.50] | [-1.14, -0.02] |
| | 146 | -0.57 [-0.62, -0.52] | [-1.19, 0.03] |
| | 101 | -0.58 [-0.62, -0.54] | [-1.13, -0.04] |
| | 104 | -0.59 [-0.64, -0.53] | [-1.18, -0.02] |
| | 122 | -0.86 [-1.04, -0.71] | [-1.34, -0.35] |
| | 144 | -0.87 [-0.91, -0.83] | [-1.38, -0.35] |

# Supplementary Material

## Supplementary Methods

### Tumor Segmentation Model Development

To develop the tumor ROI model, 265 slides were randomly sampled from the training split. Pathologist annotations were collected for multiple classes on each slide including adenocarcinoma, normal epithelium, atypical epithelium, necrosis, and an “other” category comprised of entities within tumor containing regions such as fibrosis, ulceration, large areas of stroma within tumor, and areas with evidence of treatment effect. Regions such as normal non-epithelial tissue (e.g., muscle and submucosa) were not annotated.

A sample annotated slide is provided in Supplementary Figure S3. The annotated slides were split into train, tune and test splits in a ratio of 3:1:1. After reviewing notes provided by the annotator, 21 slides were dropped either due to slide quality issues or incomplete annotations. This resulted in 149 slides for training, 51 slides for tuning and 44 slides for testing (all within the training split). A convolutional neural network based on the Inception-v3 [57] architecture with reduced parameters (depth_multipler=0.1) was trained to distinguish between adenocarcinoma and all other classes on a per patch basis. Details on model architecture and hyper-parameter tuning are in Supplementary Table S8.

### Region of Interest Mask Generation

The tumor model was used to generate binary ROI masks for all slides. Running the tumor model with a stride of 64 (at magnification 20X, 0.5µm per pixel) resulted in tumor probability heatmaps of resolution 32µm per “superpixel”. To generate binary ROI masks from the continuous tumor probability output of the tumor model, a threshold $t$ was selected to binarize the tumor model output for each patch. Next, denoising was performed by computing the connected components of positive regions and removing components with fewer than 8 superpixels. Finally, to include tumor-proximal regions in addition to tumor when training the survival model, the tumor-positive regions from the tumor model were dilated with a circular filter of radius $r$. For optimizing selection of $t$ and $r$, ROI masks were generated for three different values of the probability threshold $t$ and the dilation radius $r$ when tuning the prognostic model. The thresholds evaluated during tuning corresponded to recall of 95%, 90% and 75% on the tune split (Supplementary Table S9). The values used for the dilation radius $r$ were 0, 4 and 16 superpixels. The threshold $t$ and dilation radius $r$ were selected to optimize DLS performance on the tune split of the entire development set. During inference, we aligned the ROI masks to the output resolution of the prognostic model (patch size of 256 pixels across at 5X magnification, or 512µm). Only image patches where at least half of the patch was contained in the ROI mask were used for prognostication.

# Supplementary Figures

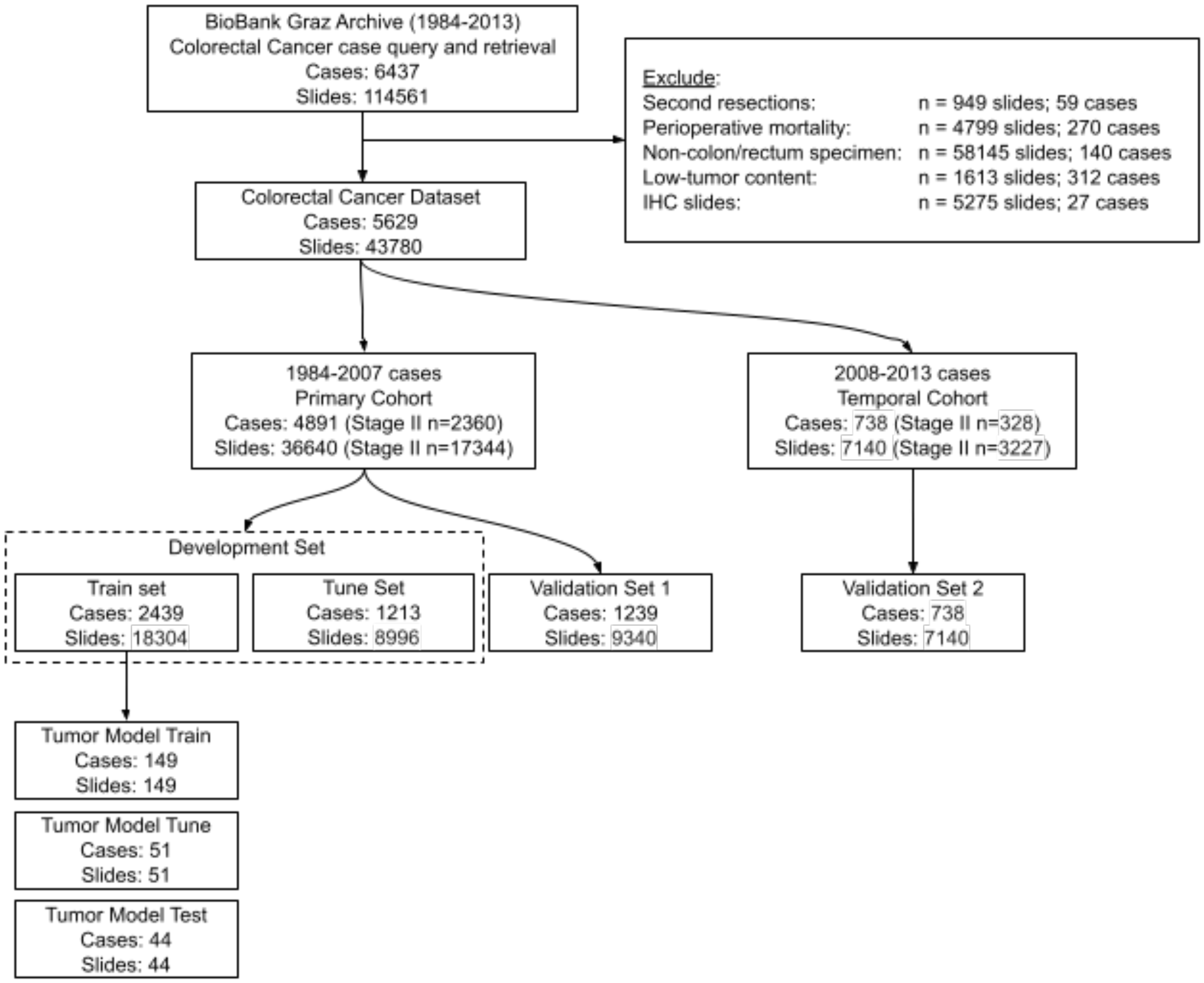

**Supplementary Figure S1. STARD diagram of the dataset curation process.**

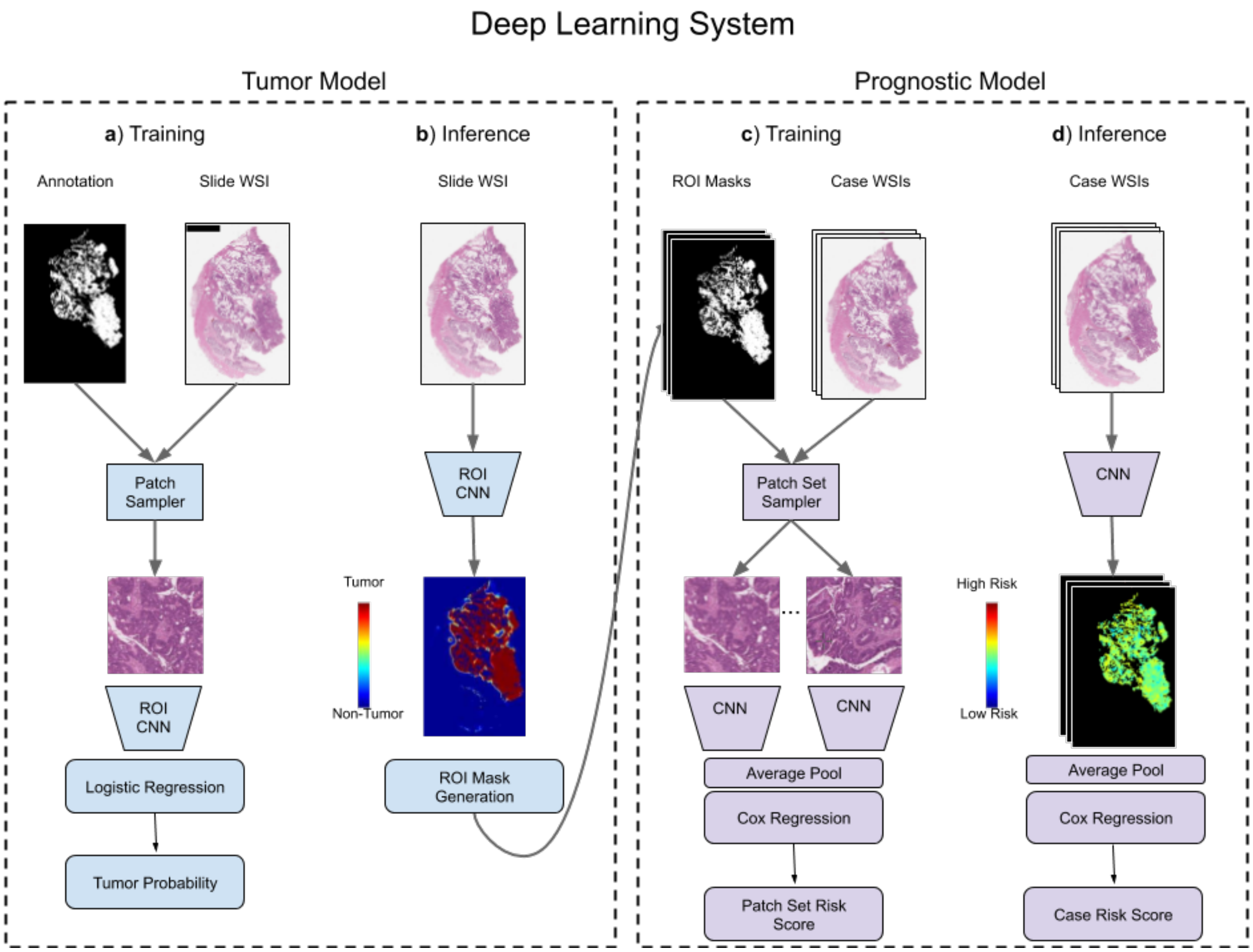

**Supplementary Figure S2. Overview of deep learning system (DLS) development**. (**a**) tumor model development: the tumor model was trained at the patch-level to identify colorectal adenocarcinoma from pixel-level pathologist annotations. (**b**) tumor model inference: the tumor model was run over all slides to produce region of interest (ROI) heatmaps that were binarized to generate ROI masks. (**c**) prognostic model development: The model was trained to predict case-level disease-specific survival. During training, a case is approximated by sampling a small number of patches from across the ROIs in a case. (**d**) prognostic model inference: at inference time, the prognostic model was run exhaustively across all ROIs to produce a case-level risk score. Scale bar indicates 5 mm. Note that the patch sampler's output image patches are shown for illustrative purposes only; the actual patch sizes will vary depending on the magnification (Supplementary Tables S8 and S10).

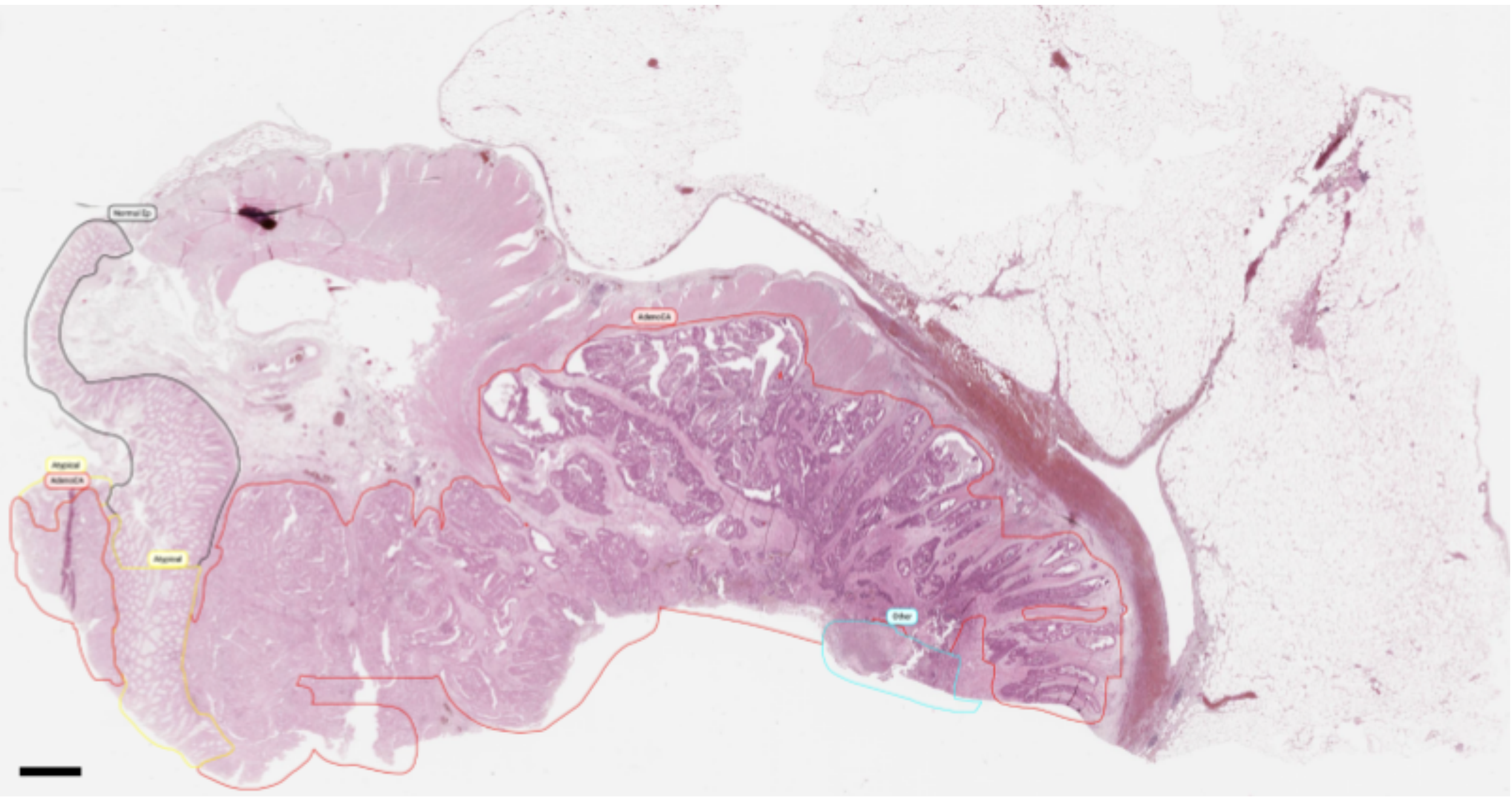

**Supplementary Figure S3. Example of slide annotations for tumor model development.** Annotations were provided for multiple types of histologies (e.g. normal epithelium, adenocarcinoma, atypical, and “other”). The model was developed to differentiate between colon adenocarcinoma and all other classes. Scale bar indicates 1 mm.

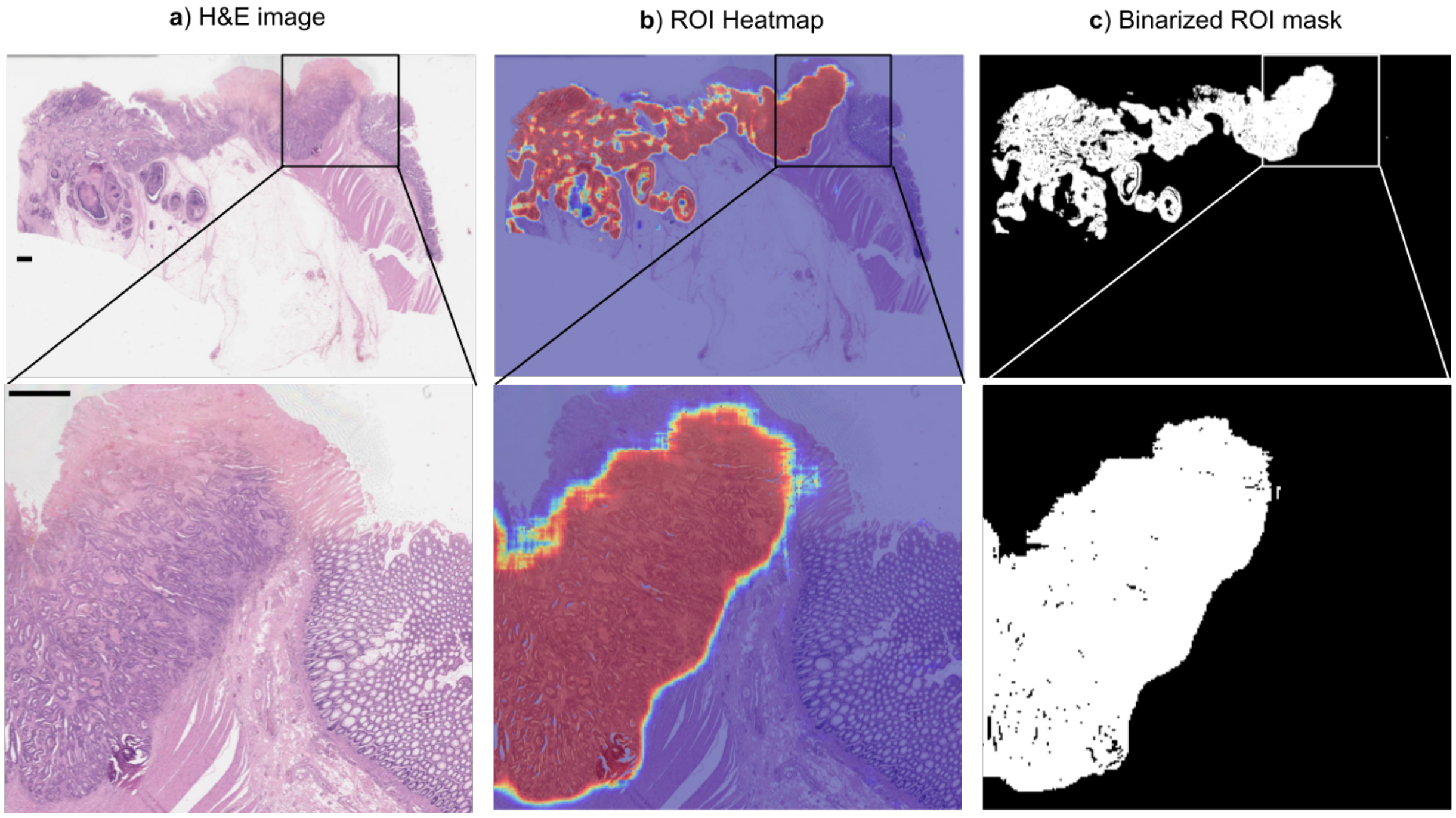

**Supplementary Figure S4. Sample tumor segmentation model predictions and derived binary ROI mask that is used to sample image patches for the prognostic model.** Scale bars indicate 1 mm.

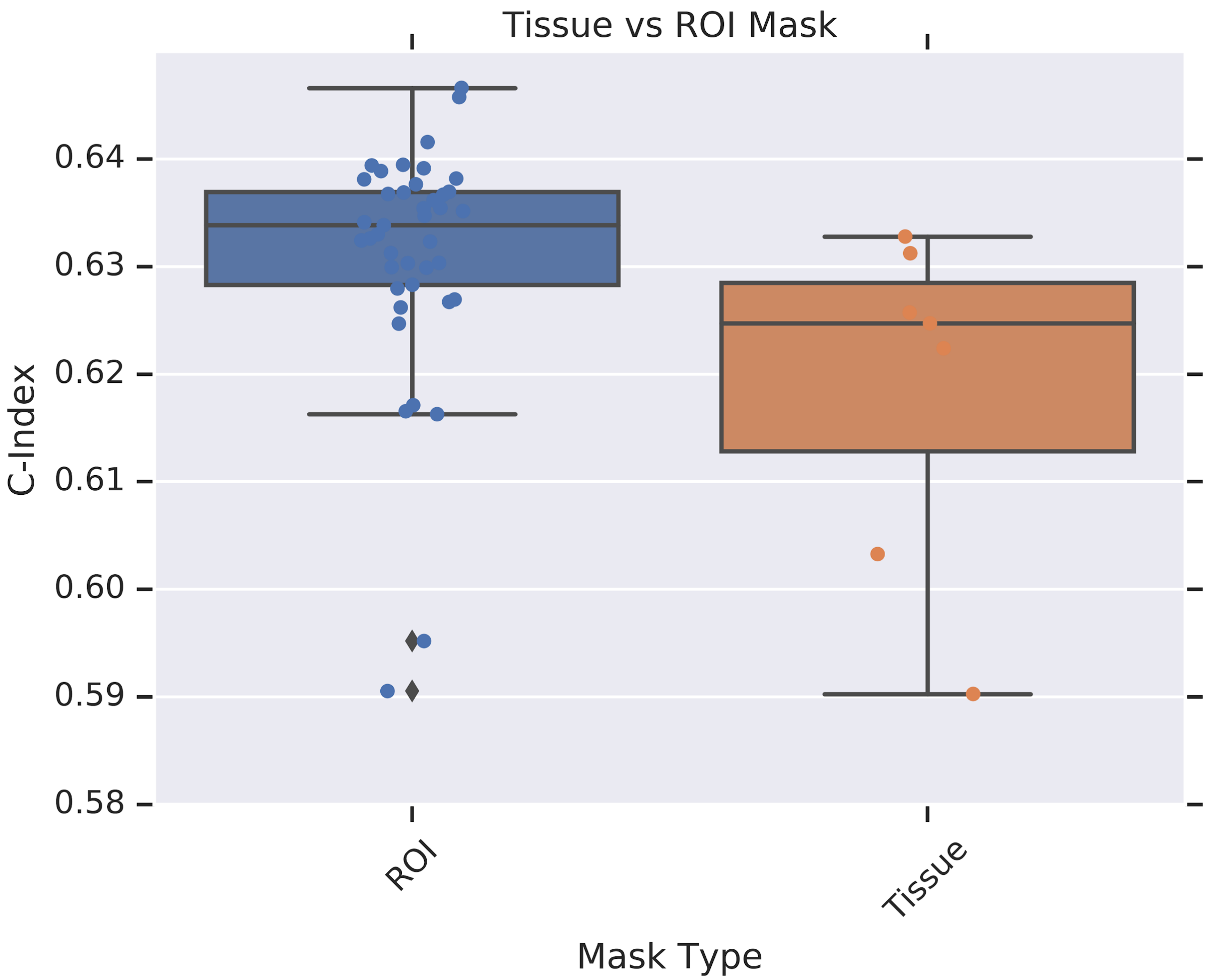

**Supplementary Figure S5. Comparison of training using the entire tissue versus on a region of interest (ROI) derived using the tumor segmentation model**. Variation in these box plots stems from different learning rates for both types of models and different mask generation parameters for the models trained on ROI masks. Models were evaluated on the tune set. Edges of boxes indicate quartiles, whiskers represent the ranges, and outliers are defined by 1.5 times the interquartile range.

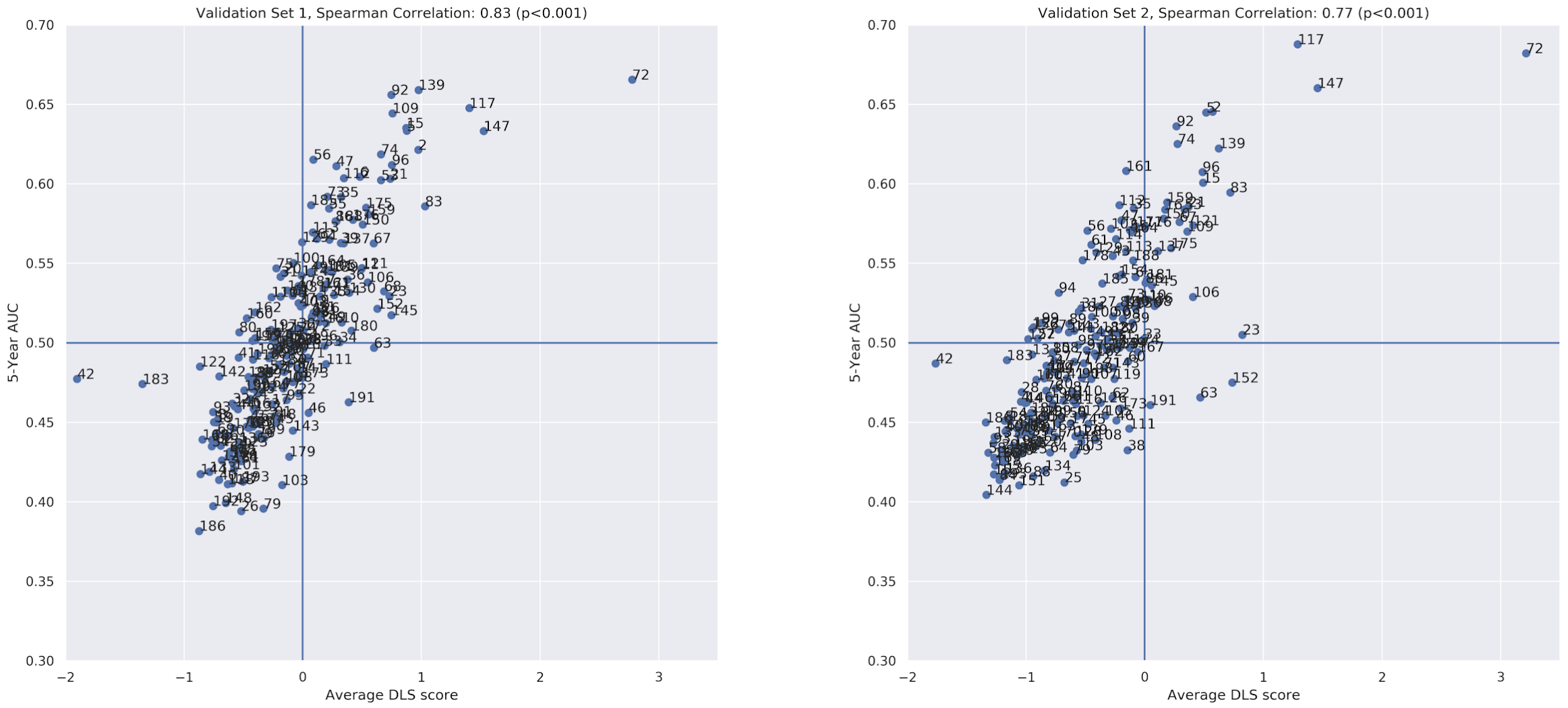

**Supplementary Figure S6. Association of each clustering-derived feature's DLS score and the 5-year AUC in both validation sets.**

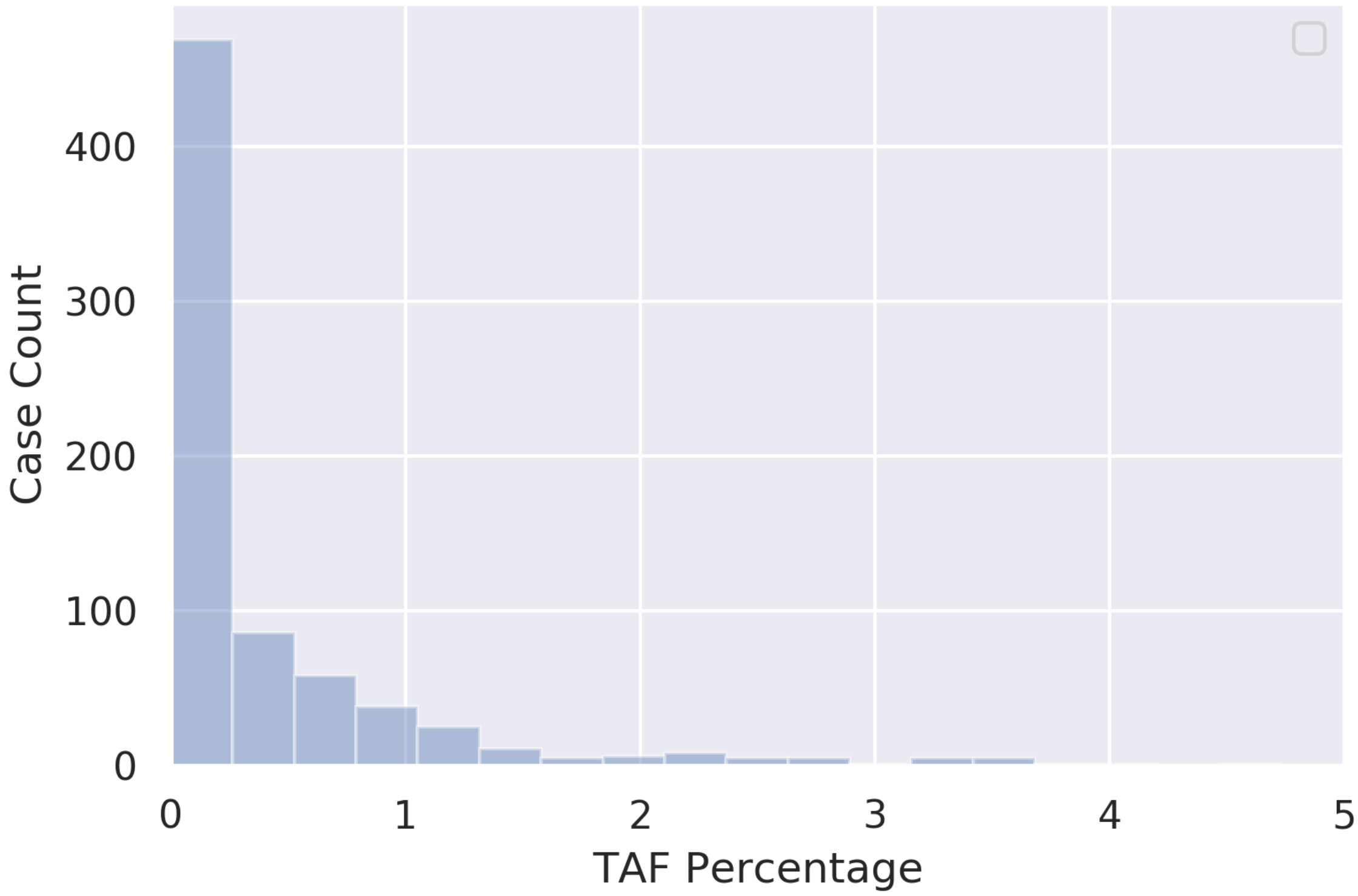

**Supplementary Figure S7. Histogram of the percentage of the region of interest that is composed of the tumor-adipose feature (TAF) in validation set 2.**

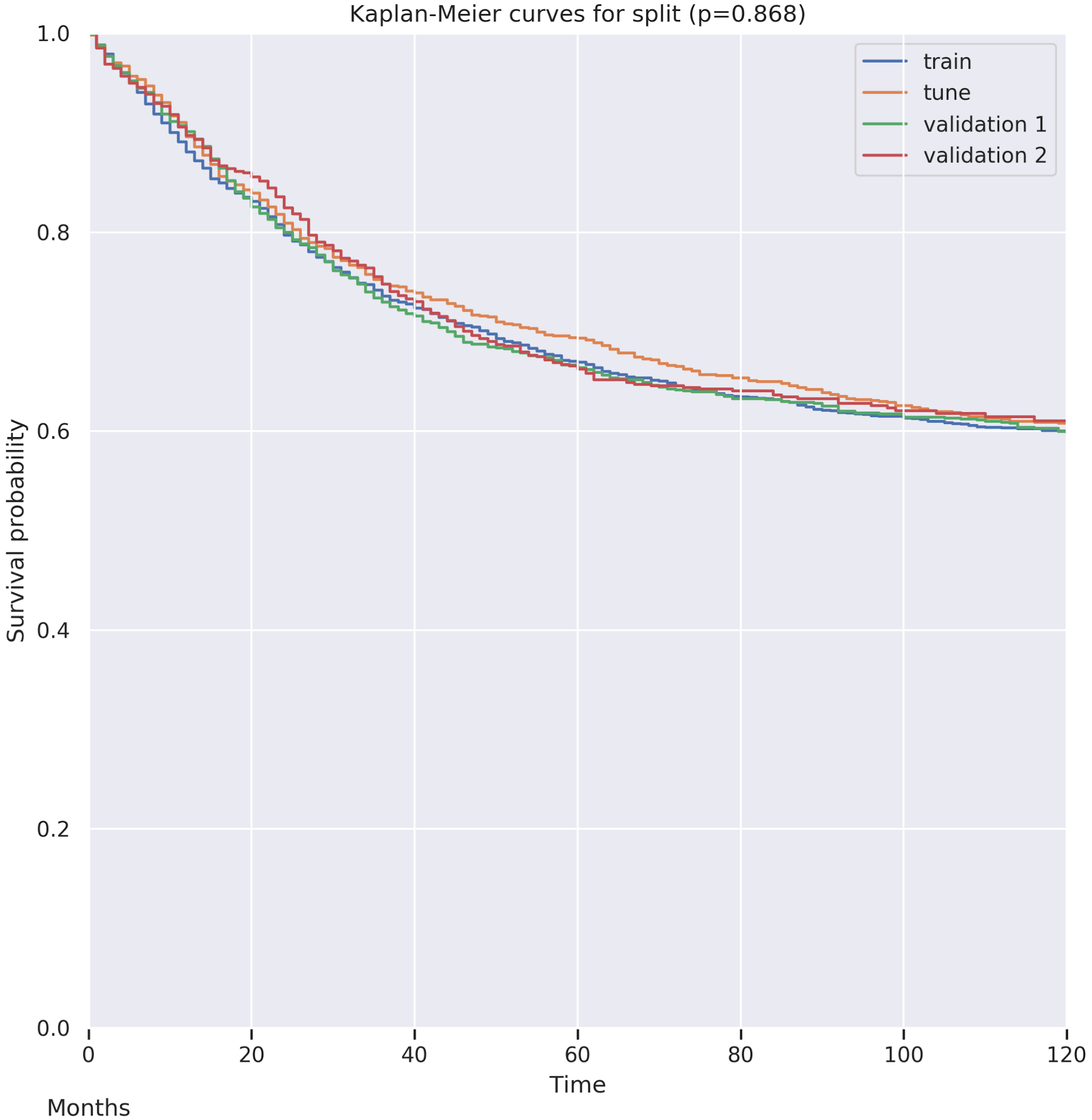

| Months | 0 | 20 | 40 | 60 | 80 | 100 | 120 |
|---|---|---|---|---|---|---|---|
| train | 2439 | 1948 | 1638 | 1440 | 1311 | 1205 | 1106 |
| tune | 1213 | 988 | 837 | 755 | 673 | 609 | 558 |
| validation 1 | 1239 | 986 | 820 | 722 | 662 | 617 | 558 |
| validation 2 | 738 | 607 | 498 | 433 | 348 | 238 | 124 |

**Supplementary Figure S8. Kaplan Meier curves for all cases in the train, tune, and validation sets.**

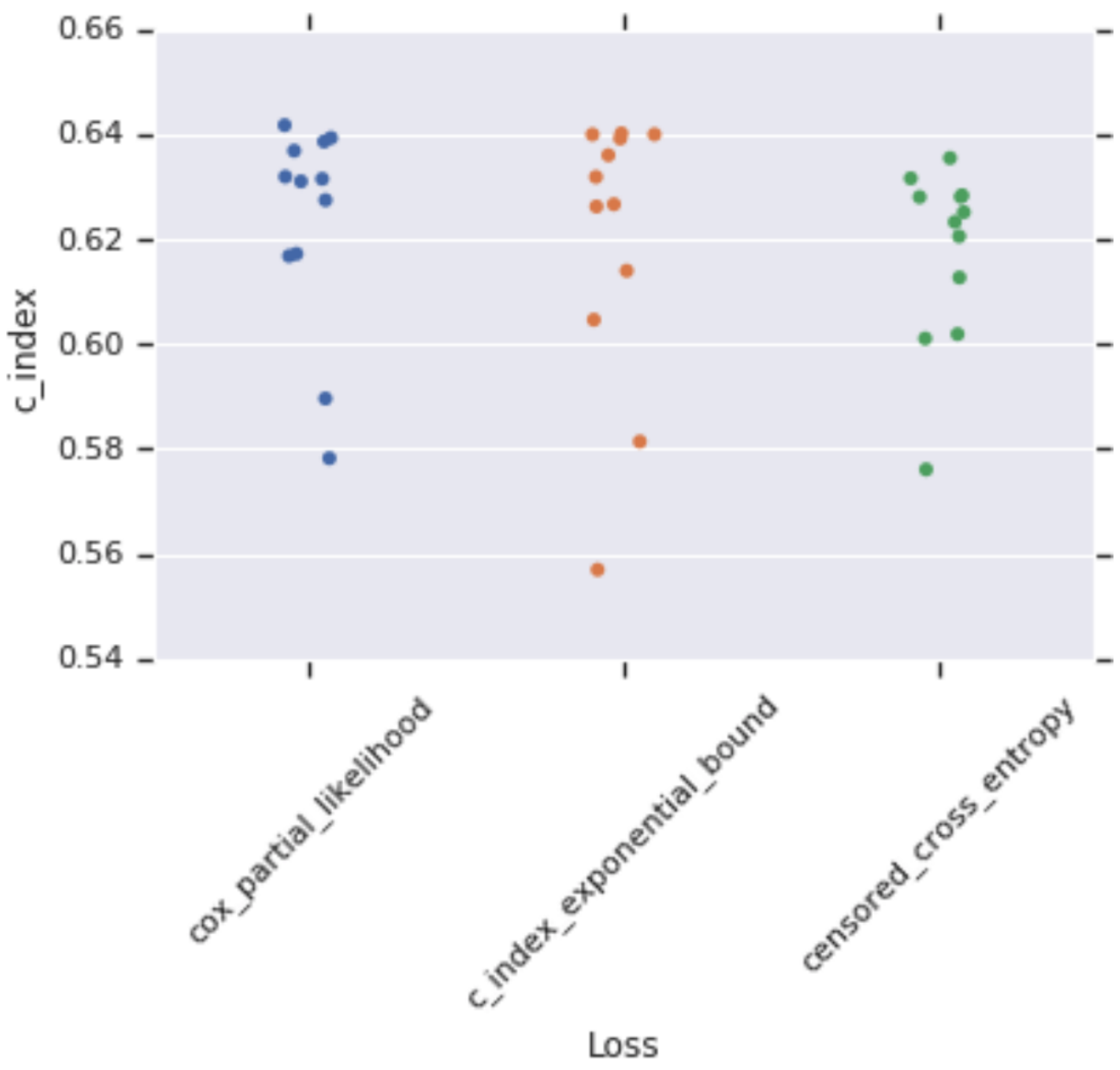

**Supplementary Figure S9. Comparison of loss functions for DLS training.** We compared three loss functions for DLS training: Cox partial likelihood, exponential lower bound on concordance index, and censored cross-entropy. For each loss function, 3 batch sizes (64, 128, 256) and 4 learning rates (10e-3, 5e-4, 10e-4, 5e-5, 10e-5) were tried. Models were evaluated on the tune set.

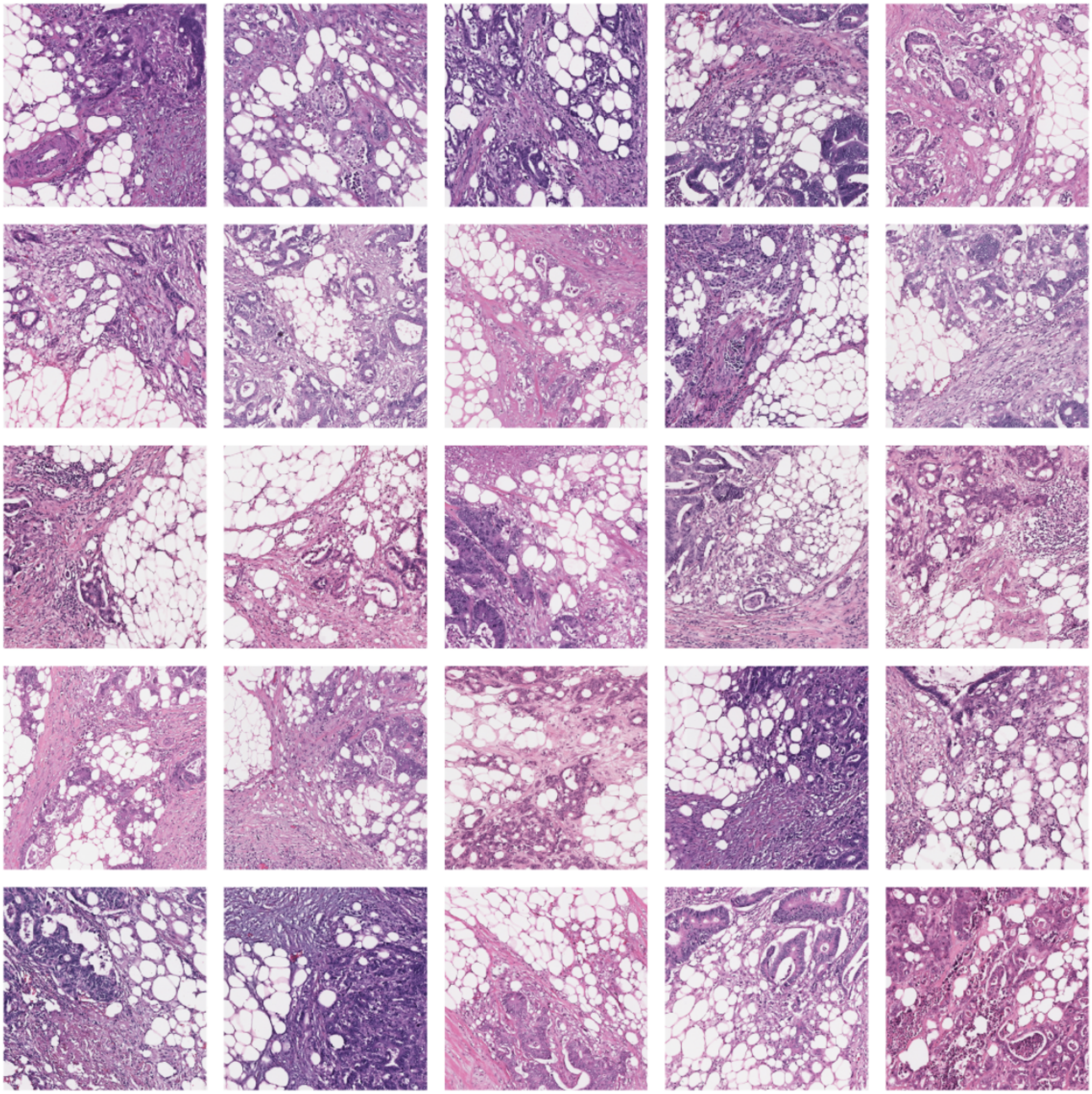

**Supplementary Figure S10. Sample patches of the TAF cluster (each from a unique case), but with the clustering centroids fit on validation set 2 and used to extract patches from validation set 1.**

# Supplementary Tables

## Supplementary Table S1. Clinical metadata distribution of the two validation sets.

| | | Stage II | | | Stage III | | |
|---|---|---|---|---|---|---|---|
| | | **Validation set 1** | **Validation set 2** | **P-value for difference** | **Validation set 1** | **Validation set 2** | **P-value for difference** |
| **T category** | **T1/T2** | 0 (0%) | 0 (0%) | N/A | 70 (11%) | 42 (10%) | 0.7083 |
| | **T3** | 546 (91%) | 270 (82%) | **0.0004** | 439 (69%) | 254 (62%) | **0.0235** |
| | **T4** | 55 (9%) | 58 (18%) | **0.0004** | 129 (20%) | 114 (28%) | **0.0055** |
| **N category** | **N0** | 601 (100%) | 328 (100%) | N/A | 0 (0%) | 0 (0%) | N/A |
| | **N1** | 0 (0%) | 0 (0%) | N/A | 361 (57%) | 245 (60%) | 0.3095 |
| | **N2** | 0 (0%) | 0 (0%) | N/A | 189 (30%) | 158 (39%) | **0.0032** |
| | **N3** | 0 (0%) | 0 (0%) | N/A | 88 (14%) | 7 (2%) | **0.0000** |
| **R category** | **R0** | 588 (98%) | 320 (98%) | 0.7907 | 606 (95%) | 392 (96%) | 0.6388 |
| | **R1** | 13 (2%) | 8 (2%) | 0.7907 | 32 (5%) | 18 (4%) | 0.6388 |
| **L category** | **L0** | 532 (89%) | 272 (83%) | **0.0231** | 501 (79%) | 274 (67%) | **0.0000** |
| | **L1** | 69 (11%) | 56 (17%) | **0.0231** | 137 (21%) | 136 (33%) | **0.0000** |
| **V category** | **V0** | 580 (97%) | 295 (90%) | **0.0004** | 583 (91%) | 312 (76%) | **0.0000** |
| | **V1** | 21 (3%) | 33 (10%) | **0.0004** | 55 (9%) | 98 (24%) | **0.0000** |
| **Tumor grade** | **G1** | 27 (4%) | 23 (7%) | 0.1264 | 16 (3%) | 17 (4%) | 0.1598 |
| | **G2** | 464 (77%) | 219 (67%) | **0.0009** | 428 (67%) | 226 (55%) | **0.0001** |
| | **G3** | 102 (17%) | 80 (24%) | **0.0089** | 188 (29%) | 155 (38%) | **0.0056** |
| | **GX** | 8 (1%) | 6 (2%) | 0.5700 | 6 (1%) | 12 (3%) | **0.0307** |
| **Self-reported sex** | **Male** | 340 (57%) | 202 (62%) | 0.1369 | 339 (53%) | 204 (50%) | 0.2861 |
| | **Female** | 261 (43%) | 126 (38%) | 0.1369 | 299 (47%) | 206 (50%) | 0.2861 |
| Age at diagnosis | **<= 59** | 117 (19%) | 43 (13%) | **0.0102** | 149 (23%) | 90 (22%) | 0.5960 |
| | **60-69** | 166 (28%) | 83 (25%) | 0.4433 | 193 (30%) | 99 (24%) | **0.0290** |
| | **70-79** | 223 (37%) | 116 (35%) | 0.5982 | 210 (33%) | 120 (29%) | 0.2120 |
| | **>= 80** | 95 (16%) | 86 (26%) | **0.0003** | 86 (13%) | 101 (25%) | **0.0000** |

**Supplementary Table S2. KM estimate of 5-year disease-specific survival in risk groups stratified by the deep learning system (DLS)**. Numbers in square brackets represent 95% confidence intervals.

| Dataset | Risk Group | Stage II | Stage III | Stage II/II |
|---|---|---|---|---|
| Validation set 1 | High (top quartile) | 63.10 [54.26, 70.70] | 34.90 [26.98, 42.91] | 45.82 [40.13, 51.32] |
| | Intermediate (middle quartiles) | 81.75 [76.56, 85.89] | 55.74 [50.02, 61.06] | 67.73 [63.59, 71.51] |
| | Low (bottom quartile) | 89.48 [83.81, 93.24] | 65.83 [57.69, 72.78] | 83.03 [78.51, 86.67] |
| Validation set 2 | High (top quartile) | 57.07 [44.05, 68.13] | 42.72 [32.25, 52.76] | 46.10 [38.28, 53.56] |
| | Intermediate (middle quartiles) | 77.76 [68.87, 84.40] | 52.82 [44.80, 60.21] | 64.83 [58.78, 70.22] |
| | Low (bottom quartile) | 85.56 [78.29, 90.54] | 73.07 [64.79, 79.70] | 80.01 [74.66, 84.35] |

**Supplementary Table S3. Univariable Cox regression on the validation sets.** Numbers indicate hazard ratio followed by 95% confidence intervals in square brackets, and p-values (from a Wald test) after the comma. *N/A because stage II only contains N0 and T3 or T4 and stage II only contains N1 by definition (American Joint Committee on Cancer, AJCC). Bold indicates statistically significant input variables ($p < 0.05$).

| **Variable** | Stage II | | Stage III | | Stage II/III | |
|---|---|---|---|---|---|---|
| | Validation set 1 | Validation set 2 | Validation set 1 | Validation set 2 | Validation set 1 | Validation set 2 |
| **DLS** | **1.64 [1.40, 1.92], <0.001** | **1.55 [1.25, 1.92], <0.001** | **1.49 [1.33, 1.67], <0.001** | **1.51 [1.32, 1.74], <0.001** | **1.72 [1.57, 1.89], <0.001** | **1.64 [1.47, 1.84], <0.001** |
| **Age** | 1.06 [0.91, 1.24], 0.446 | **1.49 [1.18, 1.87], <0.001** | **1.11 [1.01, 1.22], 0.025** | **1.24 [1.09, 1.41], 0.001** | 1.06 [0.98, 1.15], 0.121 | **1.25 [1.12, 1.40], <0.001** |
| **Sex** | | | | | | |
| **Male** | 1.0 (reference) | | | | | |
| **Female** | 0.78 [0.56, 1.07], 0.127 | 0.90 [0.57, 1.42], 0.653 | 0.79 [0.63, 0.98], 0.036 | 0.94 [0.70, 1.26], 0.682 | **0.80 [0.67, 0.97], 0.019** | 1.01 [0.79, 1.29], 0.929 |
| **Grade** | | | | | | |
| **G1** | 1.0 (reference) | | | | | |
| **G2** | 0.78 [0.38, 1.60], 0.503 | 1.67 [0.61, 4.62], 0.320 | 1.27 [0.56, 2.86], 0.563 | 3.06 [0.97, 9.65], 0.056 | 1.09 [0.64, 1.86], 0.754 | **2.36 [1.11, 5.03], 0.027** |
| **G3** | 1.17 [0.54, 2.54], 0.682 | 1.49 [0.51, 4.42], 0.467 | 1.89 [0.83, 4.31], 0.128 | **3.74 [1.18, 11.87], 0.025** | **1.81 [1.05, 3.14], 0.034** | **2.94 [1.36, 6.33], 0.006** |
| **GX** | 0.90 [0.19, 4.22], 0.889 | 2.38 [0.44, 13.00], 0.317 | 0.90 [0.18, 4.47], 0.899 | 2.92 [0.70, 12.21], 0.143 | 0.93 [0.31, 2.83], 0.902 | 2.75 [0.96, 7.84], 0.059 |
| **Lymphatic Invasion** | | | | | | |
| **L0** | 1.0 (reference) | | | | | |
| **L1** | **1.71 [1.12, 2.61], 0.012** | 1.02 [0.57, 1.81], 0.956 | 0.81 [0.61, 1.07], 0.138 | 1.23 [0.91, 1.66], 0.186 | 1.17 [0.92, 1.47], 0.199 | **1.35 [1.04, 1.75], 0.026** |
| **N-category** | | | | | | |
| **N0** | N/A* | | | | 1.0 (reference) | |
| **N1** | N/A* | | 1.0 (reference) | | **2.16 [1.73, 2.69], 0.001** | **1.73 [1.28, 2.33], 0.001** |
| **N2** | N/A* | | **1.29 [1.00, 1.65], 0.046** | **1.78 [1.33, 2.38], 0.001** | **2.78 [2.16, 3.57], 0.001** | **3.09 [2.28, 4.19], 0.001** |
| **N3** | N/A* | | 1.29 [0.93, 1.79], 0.129 | 0.70 [0.17, 2.83], 0.615 | **2.79 [2.00, 3.89], 0.001** | **1.21 [0.30, 4.91], 0.793** |
| **Margin Status** | | | | | | |
| **R0** | 1.0 (reference) | | | | | |
| **R1** | 1.19 [0.44, 3.21], 0.732 | 1.84 [0.58, 5.83], 0.301 | 1.44 [0.89, 2.31], 0.136 | 1.04 [0.51, 2.11], 0.921 | **1.56 [1.01, 2.39], 0.043** | 1.32 [0.72, 2.42], 0.365 |
| **T-category** | | | | | | |
| **T1/T2** | N/A* | | 1.0 (reference) | | | |
| **T3** | 1.0 (reference) | | **1.67 [1.09, 2.55], 0.017** | **2.81 [1.31, 6.06], 0.008** | 1.06 [0.70, 1.61], 0.770 | 2.02 [0.95, 4.32], 0.068 |
| **T4** | **1.68 [1.06, 2.66], 0.027** | **1.93 [1.16, 3.20], 0.011** | **2.37 [1.50, 3.75], 0.001** | **6.42 [2.96, 13.94], 0.001** | **1.90 [1.22, 2.97], 0.005** | **4.95 [2.30, 10.66], 0.001** |
| **Venous Invasion** | | | | | | |
| **V0** | 1.0 (reference) | | | | | |
| **V1** | 1.76 [0.90, 3.46], 0.099 | 1.43 [0.74, 2.77], 0.292 | 0.92 [0.61, 1.38], 0.671 | **1.63 [1.19, 2.25], 0.003** | 1.26 [0.89, 1.78], 0.199 | **1.83 [1.38, 2.43], 0.001** |

**Supplementary Table S4. (A) 5-year AUC for the deep learning system (DLS) and Cox regression models fit on the clinical metadata, and Cox models fit on both; (B) a similar table for the tumor-adipose feature (TAF) quantitation.** Numbers in square brackets represent 95% confidence intervals.

**A**

| Dataset | Stage | DLS | Clinical | Clinical + DLS | Delta |
|---|---|---|---|---|---|
| Validation set 1 | Stage II | 0.680 [0.631, 0.739] | 0.539 [0.485, 0.610] | 0.659 [0.612, 0.716] | 0.120 [0.076, 0.188] |
| | Stage III | 0.655 [0.617, 0.694] | 0.597 [0.550, 0.645] | 0.662 [0.631, 0.709] | 0.065 [0.026, 0.108] |
| | Stage II/III | 0.698 [0.660, 0.729] | 0.678 [0.642, 0.705] | 0.733 [0.697, 0.759] | 0.055 [0.036, 0.074] |
| Validation set 2 | Stage II | 0.663 [0.592, 0.730] | 0.610 [0.544, 0.657] | 0.695 [0.629, 0.746] | 0.085 [0.036, 0.150] |
| | Stage III | 0.655 [0.600, 0.707] | 0.664 [0.606, 0.720] | 0.686 [0.624, 0.736] | 0.022 [-0.022, 0.070] |
| | Stage II/III | 0.686 [0.638, 0.723] | 0.684 [0.639, 0.716] | 0.721 [0.688, 0.753] | 0.038 [0.006, 0.064] |

**B**

| Dataset | Stage | TAF | Clinical | Clinical + TAF | Delta |
|---|---|---|---|---|---|
| Validation set 1 | Stage II | 0.645 [0.598, 0.700] | 0.539 [0.485, 0.610] | 0.595 [0.543, 0.663] | 0.056 [0.034, 0.082] |
| | Stage III | 0.629 [0.593, 0.680] | 0.597 [0.550, 0.645] | 0.625 [0.587, 0.676] | 0.029 [0.012, 0.047] |
| | Stage II/III | 0.666 [0.634, 0.697] | 0.678 [0.642, 0.705] | 0.698 [0.664, 0.723] | 0.020 [0.010, 0.029] |
| Validation set 2 | Stage II | 0.634 [0.570, 0.697] | 0.610 [0.544, 0.657] | 0.620 [0.555, 0.661] | 0.010 [-0.016, 0.036] |
| | Stage III | 0.682 [0.638, 0.743] | 0.664 [0.606, 0.720] | 0.689 [0.630, 0.743] | 0.025 [0.004, 0.045] |
| | Stage II/III | 0.682 [0.641, 0.734] | 0.684 [0.639, 0.716] | 0.699 [0.653, 0.734] | 0.015 [0.006, 0.023] |

**Supplementary Table S5. C-index for the deep learning system (DLS) and Cox regression models fit on the clinical metadata, and Cox models fit on both**. Numbers in square brackets represent 95% confidence intervals.

| Dataset | Stage | DLS | Clinical | Clinical + DLS | Delta |
|---|---|---|---|---|---|
| Validation set 1 | Stage II | 0.651 [0.615, 0.703] | 0.535 [0.493, 0.596] | 0.634 [0.597, 0.680] | 0.099 [0.070, 0.143] |
| | Stage III | 0.626 [0.601, 0.655] | 0.576 [0.542, 0.613] | 0.626 [0.602, 0.654] | 0.050 [0.030, 0.082] |
| | Stage II/III | 0.663 [0.636, 0.686] | 0.640 [0.608, 0.664] | 0.685 [0.658, 0.704] | 0.045 [0.031, 0.060] |
| Validation set 2 | Stage II | 0.628 [0.568, 0.687] | 0.600 [0.554, 0.653] | 0.658 [0.607, 0.704] | 0.058 [0.015, 0.103] |
| | Stage III | 0.639 [0.597, 0.678] | 0.631 [0.591, 0.680] | 0.653 [0.609, 0.690] | 0.022 [-0.018, 0.060] |
| | Stage II/III | 0.660 [0.624, 0.694] | 0.661 [0.625, 0.688] | 0.689 [0.659, 0.721] | 0.028 [0.008, 0.050] |

**Supplementary Table S6. (A) 5-year AUC in T3 cases for the deep learning system (DLS) and Cox regression models fit on the clinical metadata, and Cox models fit on both**. **(B) a similar table for the tumor-adipose feature (TAF) quantitation.** Numbers in square brackets represent 95% confidence intervals.

**A**

| Dataset | Stage | DLS | Clinical | Clinical + DLS | Delta |
|---|---|---|---|---|---|
| Validation set 1 (T3 only) | Stage II | 0.677 [0.616, 0.739] | 0.537 [0.470, 0.598] | 0.657 [0.604, 0.714] | 0.121 [0.064, 0.179] |
| | Stage III | 0.639 [0.581, 0.684] | 0.563 [0.515, 0.620] | 0.654 [0.599, 0.708] | 0.091 [0.025, 0.129] |
| | Stage II/III | 0.697 [0.661, 0.739] | 0.668 [0.629, 0.694] | 0.733 [0.698, 0.770] | 0.065 [0.047, 0.087] |
| Validation set 2 (T3 only) | Stage II | 0.642 [0.567, 0.729] | 0.585 [0.502, 0.680] | 0.679 [0.596, 0.766] | 0.094 [0.037, 0.175] |
| | Stage III | 0.629 [0.559, 0.690] | 0.590 [0.515, 0.662] | 0.641 [0.561, 0.702] | 0.051 [-0.002, 0.116] |
| | Stage II/III | 0.654 [0.598, 0.701] | 0.641 [0.578, 0.702] | 0.685 [0.632, 0.732] | 0.044 [0.004, 0.080] |

**B**

| Dataset | Stage | TAF | Clinical | Clinical + TAF | Delta |
|---|---|---|---|---|---|
| Validation set 1 (T3 only) | Stage II | 0.645 [0.590, 0.691] | 0.537 [0.470, 0.598] | 0.592 [0.526, 0.651] | 0.055 [0.032, 0.092] |
| | Stage III | 0.618 [0.558, 0.675] | 0.563 [0.515, 0.620] | 0.602 [0.555, 0.656] | 0.038 [0.009, 0.059] |
| | Stage II/III | 0.668 [0.634, 0.703] | 0.668 [0.629, 0.694] | 0.692 [0.659, 0.720] | 0.025 [0.017, 0.035] |
| Validation set 2 (T3 only) | Stage II | 0.604 [0.530, 0.712] | 0.585 [0.502, 0.680] | 0.600 [0.512, 0.692] | 0.015 [-0.015, 0.056] |
| | Stage III | 0.653 [0.576, 0.714] | 0.590 [0.515, 0.662] | 0.633 [0.564, 0.709] | 0.043 [0.022, 0.070] |
| | Stage II/III | 0.649 [0.599, 0.707] | 0.641 [0.578, 0.702] | 0.666 [0.612, 0.721] | 0.025 [0.011, 0.039] |

**Supplementary Table S7. Spearman correlation between clinicopathologic features and (A) the deep learning system (DLS) or (B) automatic quantitation of the tumor-adipose feature.** P-values (from a t-test) are shown in parentheses. Cells with a p-value below 0.05 are bolded. Abbreviations for L/N/R/T/V are defined in the "Data Cohorts" section of Methods.

**A**

| Dataset | Stage | T | N | R | L | V | G | Sex | Age |
|---|---|---|---|---|---|---|---|---|---|
| Validation set 1 | Stage II | 0.07 (0.080) | N/A | -0.08 (0.057) | 0.07 (0.084) | 0.02 (0.684) | **0.13 (0.002)** | 0.0 (0.928) | **-0.09 (0.024)** |
| | Stage III | **0.27 (<0.001)** | **0.22 (<0.001)** | **0.14 (<0.001)** | -0.06 (0.141) | **0.11 (0.006)** | **0.23 (<0.001)** | 0.03 (0.421) | -0.07 (0.067) |
| | Stage II/III | **0.18 (<0.001)** | **0.36 (<0.001)** | **0.07 (0.009)** | 0.04 (0.179) | **0.10 (<0.001)** | **0.22 (<0.001)** | 0.03 (0.322) | **-0.10 (0.001)** |
| Validation set 2 | Stage II | **0.18 (0.001)** | N/A | 0.11 (0.054) | 0.09 (0.093) | **0.14 (0.010)** | **0.17 (0.003)** | 0.07 (0.183) | 0.04 (0.517) |
| | Stage III | **0.27 (<0.001)** | **0.19 (<0.001)** | **0.10 (0.038)** | **0.13 (0.008)** | **0.16 (0.001)** | **0.17 (0.001)** | -0.01 (0.791) | -0.04 (0.433) |
| | Stage II/III | **0.24 (<0.001)** | **0.34 (<0.001)** | **0.12 (0.001)** | **0.17 (<0.001)** | **0.21 (<0.001)** | **0.20 (<0.001)** | 0.06 (0.115) | -0.04 (0.339) |

**B**

| Dataset | Stage | T | N | R | L | V | G | Sex | Age |
|---|---|---|---|---|---|---|---|---|---|
| Validation set 1 | Stage II | **0.12 (0.003)** | N/A | 0.01 (0.890) | 0.04 (0.371) | -0.00 (0.986) | **0.15 (0.000)** | -0.02 (0.617) | -0.00 (0.974) |
| | Stage III | **0.36 (0.000)** | **0.13 (0.001)** | **0.12 (0.003)** | 0.02 (0.573) | **0.16 (0.000)** | **0.16 (0.000)** | -0.03 (0.413) | 0.05 (0.230) |
| | Stage II/III | **0.27 (0.000)** | **0.28 (0.000)** | **0.09 (0.002)** | **0.06 (0.024)** | **0.12 (0.000)** | **0.18 (0.000)** | -0.02 (0.513) | 0.01 (0.785) |
| Validation set 2 | Stage II | **0.17 (0.002)** | N/A | **0.16 (0.005)** | 0.03 (0.611) | **0.12 (0.025)** | 0.05 (0.384) | -0.14 (0.012) | 0.05 (0.357) |
| | Stage III | **0.46 (0.000)** | **0.17 (0.000)** | 0.08 (0.093) | **0.14 (0.005)** | **0.20 (0.000)** | 0.03 (0.591) | -0.07 (0.161) | 0.05 (0.278) |
| | Stage II/III | **0.37 (0.000)** | **0.23 (0.000)** | **0.12 (0.001)** | **0.13 (0.000)** | **0.20 (0.000)** | **0.07 (0.072)** | -0.07 (0.053) | 0.04 (0.282) |

**Supplementary Table S8. Hyperparameter search space and optimal hyperparameters for the tumor segmentation model**. We used random search (n=50 configurations) and selected the best model checkpoint based on the tuning set 5-year AUC.

| Hyperparameter | Description | Values | Optimal configuration |
|---|---|---|---|
| Batch size | Number of examples in each training batch | 64 | 64 |
| Patch size | Height and width of each image patch | 299 | 299 |
| Magnification | Image magnification at which the patches are extracted | 20X, 10X, 5X, 2.5X, 1.25X | 5X |
| Neural network architecture | Convolutional neural network architecture | InceptionV3 | InceptionV3 |
| Depth Multiplier | Multiplier on the depth of each convolution layer for downscaling the number of parameters in the default network architecture | 0.05, 0.1, 0.15, 0.2 | 0.1 |
| Loss | Loss function used for training | Softmax cross-entropy | Softmax cross-entropy |
| Optimizer | The optimization algorithm used for model training | RMSProp | RMSProp |
| L2 regularization weight | Weight of the L2 loss used for regularization | 0.001, 0.0001, 0.00001 | 0.0001 |
| Initial learning rate | Initial learning rate used for the RMSPROP optimizer; decay rate was 0.99 every 20,000 steps | 0.005, 0.0005, 0.00005 | 0.005 |
| Learning rate decay steps | Number of steps after which the learning rate is decreased by multiplying by the decay rate | 10000, 20000 | 10000 |
| Learning rate decay rate | The rate at which the learning rate is decayed after a fixed number of steps | 0.95, 0.99 | 0.99 |
| Exponential moving average decay rate | Decay rate used for taking an exponential moving average of the model weights for evaluation | None, 0.999, 0.9999 | 0.999 |
| Training steps | The number of steps for which the model is trained | 2000000 | 2000000 |
| Evaluation steps | The number of train steps after which the model is evaluated | 10000 | 10000 |

**Supplementary Table S9. Tumor segmentation model performance on its test split at three different thresholds.** Thresholds were chosen based on the recall observed on the tune split. AUC was 98.50.

| **Threshold** | **Recall** | **Precision** | **Intersection over union** |
|---|---|---|---|
| 95% tune set recall | 97.58 | 83.38 | 93.63 |
| 90% tune set recall | 93.99 | 88.58 | 94.72 |
| 75% tune set recall | 81.42 | 93.81 | 93.02 |

**Supplementary Table S10. Hyperparameter search space and optimal hyperparameters for the prognostic model**. (**A**) We used random search (n=100 configurations across the search space and selected the best model checkpoint based on the tuning set 5-year AUC. (**B**) The final DLS predictions were generated by ensembling the top 5 models.

**A**

| Hyperparameter | Description | Value |
|---|---|---|
| Batch size* | Number of examples in each training batch. | 64 |
| Patch size* | Height and width of each image patch. | 256 |
| Patch set size* | Number of patches sampled from a case to form a single training example: | 16 |
| Magnification | Image magnification at which the patches are extracted | 20X, 10X, 5X, 2.5X |
| ROI model recall | The recall for tumor detection. Recall of 100 corresponds to using a tissue mask instead of an ROI mask. | 100, 95, 90, 75 |
| ROI region dilation | The number of superpixels by which the ROI mask is dilated | 0, 4, 16 |
| Number of layers | Number of layers used in our MobileNet-based architecture | 4, 8 |
| Base depth | Depth of the first convolution layer in the network | 8, 16, 32 |
| Depth growth rate | The rate at which depth grows after each stride 2 layer. | 1.25, 1.5, 2.0 |
| Max depth | The maximum depth of any layer in the network | 64, 256 |
| Loss | Survival loss function used for training. | Cox partial likelihood |
| Optimizer | The optimization algorithm used for model training. | Adam |
| L2 regularization weight | Weight of the L2 loss used for regularization | 0.001, 0.0001, 0.00001 |
| Initial Learning rate | Initial learning rate used for the RMSPROP optimizer; decay rate was 0.99 every 20,000 steps. | 0.005, 0.0005, 0.00005 |
| Learning rate decay steps | Number of steps after which the learning rate is decreased by multiplying by the decay rate. | 10000, 20000 |
| Learning rate decay rate | The rate at which the learning rate is decayed after a fixed number of steps. | 0.95, 0.99 |
| Exponential moving average decay rate | Decay rate used for taking an exponential moving average of the model weights for evaluation. | None, 0.999, 0.9999 |
| Training steps | The number of steps for which the model is trained. | 2000000 |
| Evaluation steps | The number of train steps after which the model is evaluated. | 10000 |

* These parameters were chosen based on preliminary tuning experiments. The best values from these experiments were chosen for the full hyper-parameter tuning run described here.

**B**

| **Hyperparameter** | **Model 1** | **Model 2** | **Model 3** | **Model 4** | **Model 5** |
|---|---|---|---|---|---|
| Magnification | 5X | 5X | 5X | 5X | 5X |
| ROI model recall | 90 | 90 | 90 | 90 | 95 |
| ROI region dilation | 16 | 4 | 4 | 4 | 16 |
| Number of layers | 8 | 4 | 4 | 8 | 8 |
| Base depth | 32 | 32 | 32 | 8 | 32 |
| Depth growth rate | 1.5 | 1.25 | 1.5 | 1.25 | 1.25 |
| Max depth | 256 | 64 | 64 | 256 | 256 |
| L2 Regularization | 1e-05 | 0.001 | 1e-05 | 0.001 | 0.001 |
| Initial learning rate | 0.0005 | 0.0005 | 0.0005 | 0.0005 | 5e-05 |
| Learning rate decay steps | 10000 | 10000 | 10000 | 20000 | 20000 |
| Learning rate decay rate | 0.95 | 0.95 | 0.95 | 0.95 | 0.95 |
| Exponential moving average decay rate | 0.9999 | 0.9999 | 0.9999 | N/A | 0.999 |
| Training step | 1381426 | 1403469 | 1907329 | 1714445 | 1259927 |

**Supplementary Table S11. REMARK checklist for reporting.**

| Item to be reported | | Location |
|---|---|---|
| **INTRODUCTION** | | |
| 1 | State the marker examined, the study objectives, and any pre-specified hypotheses. | Last paragraph of introduction |
| **MATERIALS AND METHODS** | | |
| *Patients* | | |
| 2 | Describe the characteristics (e.g., disease stage or co-morbidities) of the study patients, including their source and inclusion and exclusion criteria. | "Data Cohorts" section |
| 3 | Describe treatments received and how chosen (e.g., randomized or rule-based). | "Data Cohorts" section |
| *Specimen characteristics* | | |
| 4 | Describe type of biological material used (including control samples) and methods of preservation and storage. | "Data Cohorts" section |
| *Assay methods* | | |
| 5 | Specify the assay method used and provide (or reference) a detailed protocol, including specific reagents or kits used, quality control procedures, reproducibility assessments, quantitation methods, and scoring and reporting protocols. Specify whether and how assays were performed blinded to the study endpoint. | "Data Cohorts" and "Prognostic Model Neural Network Architecture and Survival Loss" sections |
| *Study design* | | |
| 6 | State the method of case selection, including whether prospective or retrospective and whether stratification or matching (e.g., by stage of disease or age) was used. Specify the time period from which cases were taken, the end of the follow-up period, and the median follow-up time. | "Data Cohorts" section |
| 7 | Precisely define all clinical endpoints examined. | "Data Cohorts" section |
| 8 | List all candidate variables initially examined or considered for inclusion in models. | "Data Cohorts" and "DLS Association with Clinicopathologic Features" section, Table 4a |
| 9 | Give rationale for sample size; if the study was designed to detect a specified effect size, give the target power and effect size. | "Data Cohorts" section |
| *Statistical analysis methods* | | |
| 10 | Specify all statistical methods, including details of any variable selection procedures and other model-building issues, how model assumptions were verified, and how missing data were handled. | "Tumor Segmentation Model", "Prognostic Model Neural Network Architecture and Survival Loss", and "Understanding DLS Predictions" sections |
| 11 | Clarify how marker values were handled in the analyses; if relevant, describe methods used for cutpoint determination. | "Evaluating DLS Performance" section |
| **RESULTS** | | |
| *Data* | | |
| 12 | Describe the flow of patients through the study, including the number of patients included in each stage of the analysis (a diagram may be helpful) and reasons for dropout. Specifically, both overall and for each subgroup extensively examined report the numbers of patients and the number of events. | Table 1, Supplementary Figure S1 |
| 13 | Report distributions of basic demographic characteristics (at least age and sex), standard (disease-specific) prognostic variables, and tumor marker, including numbers of missing values. | Supplementary Table S1 |
| *Analysis and presentation* | | |
| 14 | Show the relation of the marker to standard prognostic variables. | Supplementary Table S4 and S7 |
| 15 | Present univariable analyses showing the relation between the marker and outcome, with the estimated effect (e.g., hazard ratio and survival probability). Preferably provide similar analyses for all other variables being analyzed. For the effect of a tumor marker on a time-to-event outcome, a Kaplan-Meier plot is recommended. | P5<br>Supplementary Table S3 |
| 16 | For key multivariable analyses, report estimated effects (e.g., hazard ratio) with confidence intervals for the marker and, at least for the final model, all other variables in the model. | Table 3 |
| 17 | Among reported results, provide estimated effects with confidence intervals from an analysis in which the marker and standard prognostic variables are included, regardless of their statistical significance. | Table 3, Supplementary Tables S4 and S5 |
| 18 | If done, report results of further investigations, such as checking assumptions, sensitivity analyses, and internal validation. | Tables 4,5 |
| **DISCUSSION** | | |
| 19 | Interpret the results in the context of the pre-specified hypotheses and other relevant studies; include a discussion of limitations of the study. | Throughout Discussion |
| 20 | Discuss implications for future research and clinical value. | Throughout Discussion |